\documentclass{article}

\usepackage[preprint]{neurips_2025}

\usepackage[utf8]{inputenc} 
\usepackage[T1]{fontenc}    
\usepackage{times}          

\usepackage{hyperref}       
\usepackage{url}            

\usepackage{booktabs}       
\usepackage[table,xcdraw]{xcolor} 

\usepackage{amsmath}
\usepackage{amssymb}
\usepackage{amsfonts}       
\usepackage{nicefrac}       

\usepackage{graphicx}
\usepackage{multirow}

\usepackage{microtype}      

\usepackage{xcolor}         

\usepackage{graphicx}
\usepackage{pdfpages}
\usepackage{subcaption}

\usepackage{algorithm}
\usepackage{algpseudocode}
\usepackage{algorithmicx}

\title{A Novel Benchmark and Dataset for Efficient 3D Gaussian Splatting with Gaussian Point Cloud Compression}
\author{Kangli Wang$^1$ \quad Shihao Li$^1$ \quad Qianxi Yi$^{1,2}$ \quad Wei Gao$^{1,2}$ \thanks{Corresponding Author: Wei Gao } \\
$^1$ Guangdong Provincial Key Laboratory of Ultra High Definition Immersive Media Technology, \\
Shenzhen Graduate School, Peking University, Shenzhen, China \\
$^2$ PengCheng Laboratory, Shenzhen, China \\
\{kangliwang, shihaoli\}@stu.pku.edu.cn \quad  yqianxi128@gmail.com \quad gaowei262@pku.edu.cn
\\
\\
\centerline{\qquad \textbf{\color{magenta} Project Website}: \url{https://gauspcc.github.io}} 
}

\begin{document}

\maketitle
\begin{figure}[h]
    \centering
    \vspace{-0.2cm}
    \includegraphics[width=1.0\linewidth]{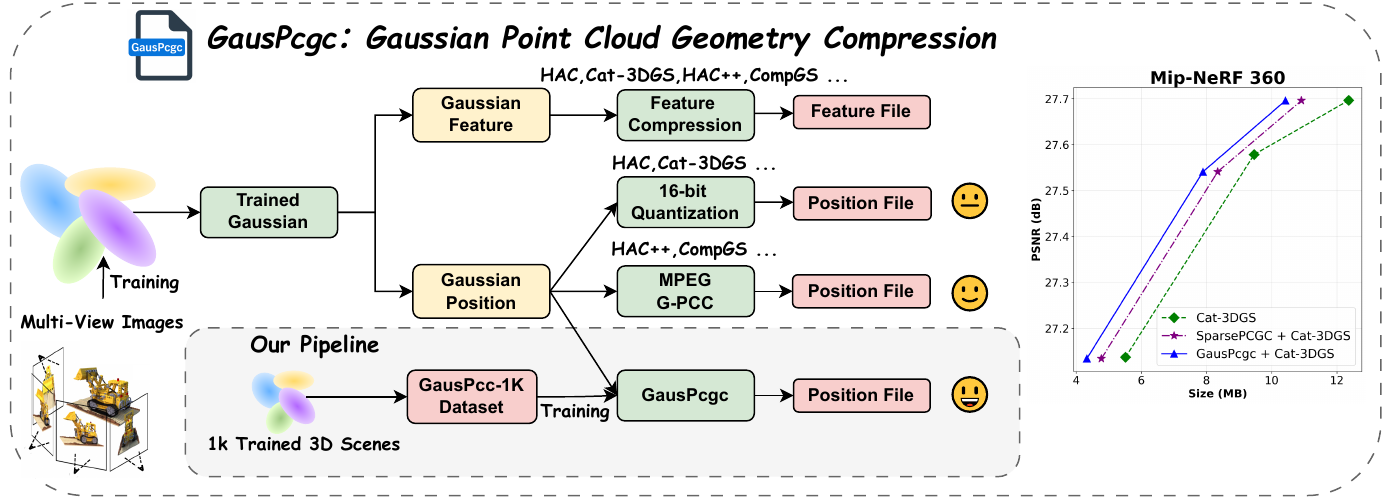}
     \caption{
     Overview of our GausPcc-1K dataset and GausPcgc framework. Existing methods neglect Gaussian positions or use suboptimal G-PCC resulting in bitstream inefficiency. GausPcgc trained on our specialized dataset achieves superior inference speed and compression rates.
     }
    \label{fig:overview}
\end{figure}

\begin{abstract}
Recently, immersive media and autonomous driving applications have significantly advanced through 3D Gaussian Splatting (3DGS), which offers high-fidelity rendering and computational efficiency. Despite these advantages, 3DGS as a display-oriented representation requires substantial storage due to its numerous Gaussian attributes. Current compression methods have shown promising results but typically neglect the compression of Gaussian spatial positions, creating unnecessary bitstream overhead.
We conceptualize Gaussian primitives as point clouds and propose leveraging point cloud compression techniques for more effective storage. AI-based point cloud compression demonstrates superior performance and faster inference compared to MPEG Geometry-based Point Cloud Compression (G-PCC). However, direct application of existing models to Gaussian compression may yield suboptimal results, as Gaussian point clouds tend to exhibit globally sparse yet locally dense geometric distributions that differ from conventional point cloud characteristics.
To address these challenges, we introduce \textbf{\textit{GausPcgc}} for Gaussian point cloud geometry compression along with a specialized training dataset \textbf{\textit{GausPcc-1K}}. Our work pioneers the integration of AI-based point cloud compression into Gaussian compression pipelines, achieving superior compression ratios. The framework complements existing Gaussian compression methods while delivering significant performance improvements. All code, data, and pre-trained models will be publicly released to facilitate further research advances in this field.

\end{abstract}

\section{Introduction}
Novel view synthesis has made significant advances in recent years. Neural Radiance Fields (NeRF) \cite{mildenhall2021nerf} implicitly encode 3D scenes, but suffer from slow training and rendering due to extensive ray point sampling. Despite various optimization efforts \cite{chen2022tensorf,kplanes}, these approaches remain computationally intensive. Recently, 3D Gaussian Splatting (3DGS) \cite{3dgs} has emerged as a new paradigm for 3D scene representation, directly modeling scenes through explicit 3D Gaussian distributions. These Gaussians feature learnable parameters that can be directly splatted onto 2D image planes, enabling fast differentiable rendering. 3DGS effectively addresses NeRF's limitations in training and rendering speed, offering broad applications in immersive media.

However, 3DGS requires substantial storage capacity to represent 3D scenes due to the numerous learnable parameters attached to each Gaussian primitive. For instance, scenes from the Mip-NeRF 360 \cite{mildenhall2021nerf} dataset demand approximately 700MB of storage per dataset. This enormous storage requirement significantly constrains 3DGS applications, driving research into Gaussian compression techniques \cite{3dgszip}. Some studies \cite{navaneet2024compgs,lee2024compact,wang2024end,xie2024mesongs} treat Gaussian attributes as feature sets and apply pruning and vector quantization to preserve only the most essential values. Nevertheless, these approaches fail to exploit spatial redundancies among Gaussian distributions, resulting in suboptimal compression efficiency. Recently, various methods \cite{chen2024hac,chen2025hac++,wang2024contextgs,ma2025enhancing,liu2024compgs,zhancat,zhan2025cat,liu2025compgs++,xie2024sizegs,wang2025tc,tang2025neuralgs,liu2024hemgs,chen2025pcgs} have focused on leveraging structural relationships between Gaussian primitives to reduce compression redundancy. While these approaches have advanced compression efficiency, most neglect the processing of Gaussian geometric positions, leading to unnecessary bitstream consumption.

Therefore, we \textbf{conceptualize Gaussian primitives as Gaussian point clouds} and propose applying point cloud geometry compression techniques to Gaussian positions, thereby enhancing the overall efficiency of Gaussian compression. However, directly transferring existing point cloud compression schemes \cite{unipcgc,sparsepcgc,you2025reno,fu2022octattention,song2023efficient,zhang2025adadpcc} to Gaussian positions compression is challenging, primarily because Gaussian geometric distributions differ significantly from conventional point clouds. Existing point cloud datasets \cite{dataset8i,kitti,dai2017scannet} feature either dense or sparse point distributions, and current point cloud compression algorithms are designed accordingly. In contrast, Gaussian geometric distributions are characterized by global sparsity with local density, which severely diminishes point cloud compression efficiency. To address this challenge, we propose establishing a new benchmark for Gaussian point cloud geometry compression through our \textbf{\textit{GausPcc-1K}} dataset and \textbf{\textit{GausPcgc}} compression method. This approach bridges the gap between traditional point cloud compression and 3DGS compression, significantly enhancing the efficiency of 3DGS compression. Specifically, to address current dataset distribution inconsistencies, we train 1,000 3DGS scenes and extracted their geometry and attributes information to create a Gaussian point cloud dataset. Using this new training dataset, we evaluate existing point cloud geometry compression algorithms and design a new compression scheme. Our designed approach seamlessly integrates into mainstream Gaussian compression frameworks and substantially reduces overall bitstream requirements. Our contributions can be summarized as:
\begin{itemize}
\item We introduce \textit{GausPcgc}, a novel benchmark and framework for Gaussian point cloud geometry compression with \textit{GausPcc-1K} as the first specialized dataset in this domain.
\item \textit{GausPcgc} pioneers the integration of AI-based point cloud compression into mainstream frameworks, enhancing Gaussian compression efficiency with minimal computational overhead. This represents the first such exploration in the field.
\item To achieve an optimal trade-off between compression efficiency and encoding complexity, we incorporate non-uniform grouping schemes and neighbor prior information in \textit{GausPcgc}, enhancing coding performance with minimal computational overhead.
\item Experimental results show our proposed \textit{GausPcgc} delivers an 8.2\% compression ratio gain over the latest G-PCC v23 for Gaussian point cloud compression. When integrated into mainstream frameworks, \textit{GausPcgc} achieves superior performance.
\end{itemize}

\section{Related works}
\subsection{3D Gaussian Compression}
Thanks to its powerful novel view synthesis capabilities, 3D Gaussian Splatting \cite{3dgs} (3DGS) has gained widespread adoption in fields such as 3D reconstruction and immersive media. However, representing a single 3D scene using 3DGS typically requires tens to hundreds of megabytes of storage, substantially increasing computational burden. Researchers have begun exploring efficient compression methods for 3DGS, primarily focusing on two approaches: eigenvalue-based compression \cite{navaneet2024compgs,lee2024compact,wang2024end,xie2024mesongs} and structured representation-based compression \cite{chen2024hac,chen2025hac++,wang2024contextgs,ma2025enhancing,liu2024compgs,zhancat,zhan2025cat,liu2025compgs++,xie2024sizegs,wang2025tc,tang2025neuralgs,liu2024hemgs,chen2025pcgs}. The eigenvalue-based approach treats 3DGS attributes as feature vectors and efficiently represents them through codebook techniques \cite{gray1984vector}. Meanwhile, structured representation-based compression, exemplified by Scaffold-GS \cite{lu2024scaffold}, structures 3D scenes by leveraging position-feature correlations for efficient context modeling to achieve compression. Nevertheless, most existing methods either neglect compressing the geometry coordinates of Gaussians entirely or rely solely on \cite{chen2025hac++,chen2025pcgs,liu2024compgs,liu2025compgs++} MPEG Geometry Point Cloud Compression (G-PCC) techniques, highlighting a significant opportunity for advancement in this domain.
\subsection{Point Cloud Compression}
Point cloud compression is divided into geometry compression and attribute compression, where geometry compression refers to the compression of the positions of point clouds, which is the focus of this article. SparsePCGC \cite{sparsepcgc}, UniPCGC \cite{unipcgc} and RENO \cite{you2025reno} represent point clouds in a voxelized manner and efficiently compress them by predicting the occupancy probability of voxels. Octattention \cite{fu2022octattention} and EHEM \cite{song2023efficient} organize point clouds in an octree manner and use attention to mine large-scale contexts for efficient probability prediction. We bridge point cloud and 3DGS compression by establishing a new compression benchmark. To our knowledge, this presents the first generalizable geometric compressor trained on large-scale 3D scenes.
\begin{figure}
    \centering
    \includegraphics[width=0.8\linewidth]{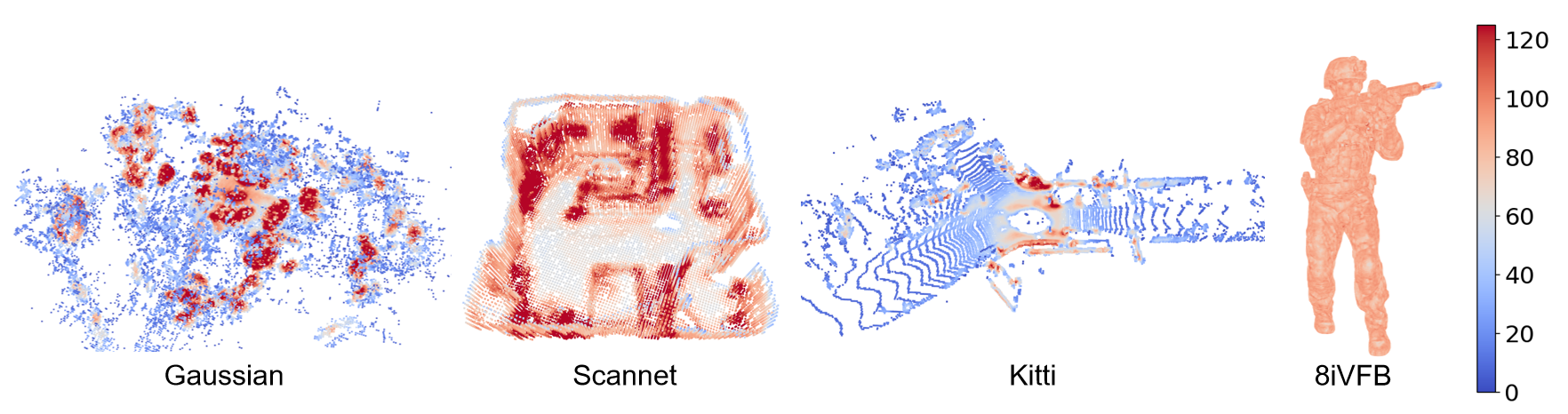}
    \caption{Visualization of local density. We count neighbors within a 5×5×5 vicinity and render the results using a color gradient. }
    \label{fig:pc_vis}
    \vspace{-8pt}
\end{figure}
\section{Preliminary}
\textbf{3DGS} \cite{3dgs} represents 3D scenes using numerous 3D primitives that are optimized through differentiable rendering. Each Gaussian primitive is precisely defined by the following formulation:
\begin{equation}
G(\boldsymbol{x})=\exp \left(-\frac{1}{2}(\boldsymbol{x}-\boldsymbol{\mu})^{\top} \boldsymbol{\Sigma}^{-1}(\boldsymbol{x}-\boldsymbol{\mu})\right),
\end{equation}
where $\boldsymbol{\mu}$ and $\boldsymbol{\Sigma}$ represent the mean and covariance of the Gaussian respectively. The covariance matrix $\boldsymbol{\Sigma}=\mathbf{R S S}^T \mathbf{R}^T$ is decomposed into scaling matrix $\mathbf{S}$ and rotation matrix $\mathbf{R}$. Each Gaussian also contains opacity $\alpha$ and color features $\boldsymbol{c}$, which serve as learnable parameters optimized to accurately represent the 3D scene.

\textbf{Scaffold-GS} \cite{lu2024scaffold} introduces anchors to capture local Gaussian similarities, significantly reducing inter-Gaussian redundancy. These anchors constitute voxelized points characterized by features $\mathbf{f} \in \mathbb{R}^{32}$, positions $\mathbf{x} \in \mathbb{R}^{3}$, scaling factors $\mathbf{I} \in \mathbb{R}^{3}$, and learnable offsets $\mathbf{O} \in \mathbb{R}^{k \times 3}$. Specifically, for an anchor point $p$, its associated mean is calculated as:
\begin{equation}
\left\{\boldsymbol{\mu}^0, \ldots, \boldsymbol{\mu}^{k-1}\right\}_p=\mathbf{x}_p+\left\{\mathbf{O}^0, \ldots, \mathbf{O}^{k-1}\right\}_p \cdot \mathbf{I}_p,
\end{equation}
where $x_p$ represents the coordinate of anchor point $p$. Given camera position $x_c$, the view-dependent properties are calculated as:
\begin{equation}
\left\{\mathbf{c}^p, \mathbf{r}^p, \mathbf{s}^p, \alpha^p\right\}=F\left(\mathbf{f}_p, \boldsymbol{\sigma}_{pc}, \overrightarrow{\mathbf{d}}_{pc}\right),
\end{equation}
where $\boldsymbol{\sigma}_{pc}=\left\|\mathbf{x}_p-\mathbf{x}_c\right\|_2, \overrightarrow{\mathbf{d}}_{pc}=\frac{\mathbf{x}_p-\mathbf{x}_c}{\left\|\mathbf{x}_p-\mathbf{x}_c\right\|_2}$, $F(\cdot)$ is the MLP network used to predict the Gaussian parameters related to the view.

\textbf{3D Point Clouds} are unordered collections of spatial points $\mathcal{P} = \{\mathbf{p}_i\}_{i=1}^N$ where $\mathbf{p}_i \in \mathbb{R}^3$. Efficient processing relies on octrees and voxel grids as hierarchical structures.
``Scale'' ($s$) denotes voxelized resolution, where scale $s=n$ discretizes space into $2^n \times 2^n \times 2^n$ voxels.
In compression, octree methods \cite{fu2022octattention,song2023efficient} offer superior efficiency despite computational demands. Sparse tensor approaches \cite{sparsepcgc,unipcgc} enable faster processing but lack efficiency for sparse distributions where $|\mathcal{P}| \ll 2^{3s}$. Recent work \cite{you2025reno} combines these approaches for optimal performance.

\section{Methodology}
\label{sec:method}
\subsection{Compression Motivation}
\label{sub: Compression Motivation}
\textbf{Why Compression.}
First, we explain why compressing geometric positions specifically is necessary for efficient 3DGS representation.
Existing works predominantly focus on feature-based or structure-based compression. Particularly for structured compression, most use Scaffold-GS \cite{lu2024scaffold} as a baseline and achieve compression by exploiting correlations between features. However, most existing methods \cite{chen2024hac,zhancat} merely quantize geometric positions with 16-bit precision without further compression. This approach requires $\text{bpp}_{\text{position}} = D \times b = 3 \times 16 = 48$ \textbf{bits per point (bpp)} solely for position encoding, substantially increasing the overall bitstream size.
For example, when HAC \cite{chen2024hac} compresses the bicycle scene from the Mip-NeRF 360 dataset \cite{barron2022mipnerf360}, position data accounts for 4.94MB of the 27.53MB total size ($S_{\text{position}}/S_{\text{total}} \approx 0.18$). This significant proportion demonstrates the considerable potential for developing specialized Gaussian point cloud geometric position compression methods to achieve more efficient representations without sacrificing visual quality.

\begin{figure}
    \centering
    \includegraphics[width=0.8\linewidth]{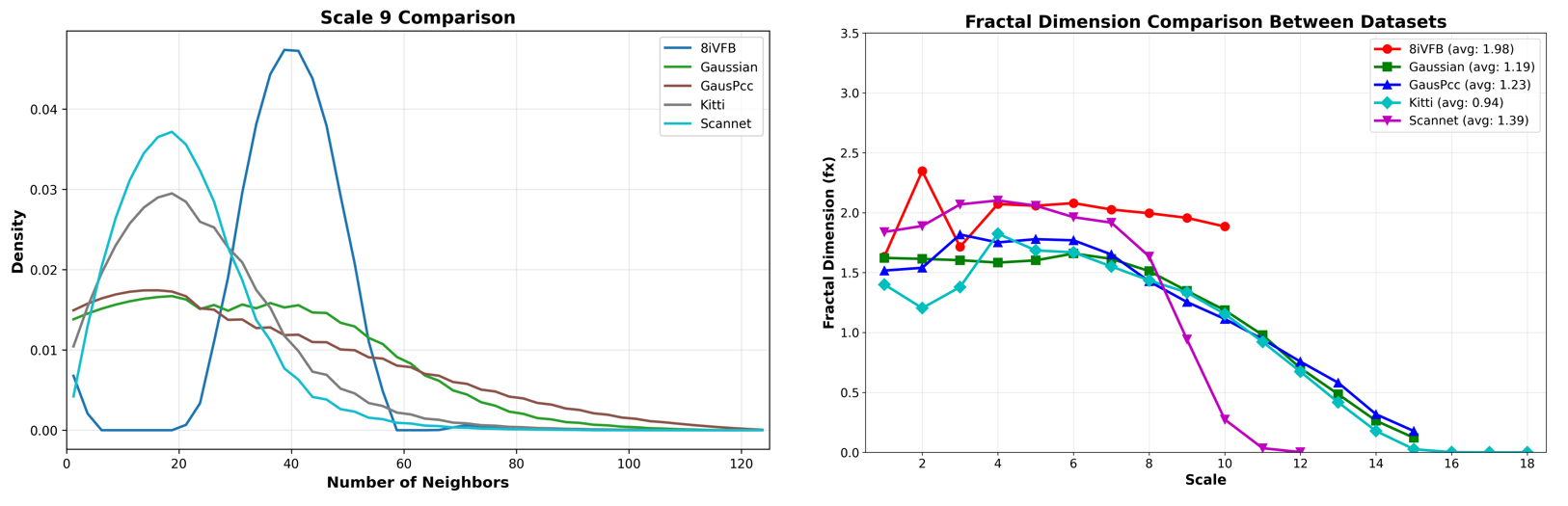}
    \caption{Comparative analysis of local density and fractal dimension across different datasets. }
    \label{fig:pc_ld_fd}
    \vspace{-8pt}
\end{figure}
\textbf{Why AI-based Compression.}
Above, we discuss the importance of compressing geometric positions for Gaussian point clouds, with recent works \cite{chen2025hac++,liu2024compgs} initiating explorations of MPEG G-PCC \cite{gpcc} compression tools. Our objective is to leverage AI-based point cloud compression methodologies to achieve enhanced compression efficiency. Contemporary AI-based approaches demonstrate substantial advantages over traditional methods in both performance metrics and computational efficiency. Quantitatively, RENO \cite{you2025reno} achieves a $\Delta BD\text{-}Rate \approx 20\%$ improvement compared to G-PCC v23 on Kitti dataset while reducing encoding time by a factor of $T_{\text{G-PCC}}/T_{\text{RENO}} \approx 15$. Similarly, on 8iVFB dataset, UniPCGC \cite{unipcgc} and SparsePCGC \cite{sparsepcgc} provide performance gains of $\Delta BD\text{-}Rate > 90\%$ with accelerated encoding processes. These empirical results strongly motivate the application of AI-based techniques to Gaussian point cloud compression.

\subsection{GausPcc-1K Dataset}
\label{sub: GausPcc-1K Dataset}
After identifying the need for Gaussian compression and advantages of AI-based methods, we apply these approaches to Gaussian data. We analyze density distributions, establish the need for a specialized dataset, detail GausPcc-1K's creation, and examine its distinctive properties.
\begin{figure}
    \centering
    \includegraphics[width=0.9\linewidth]{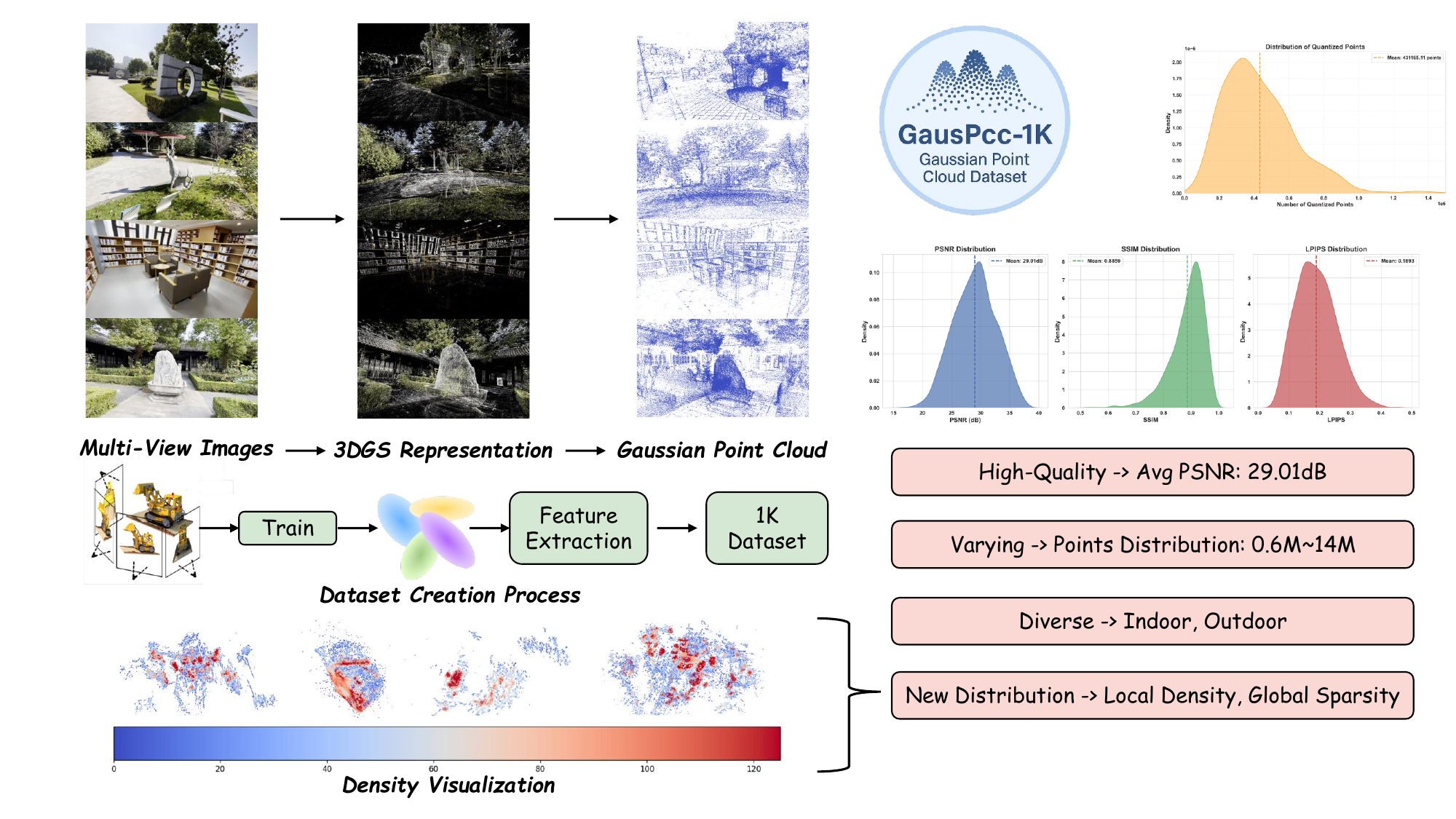}
    \caption{Introduction of the proposed GausPcc-1K Dataset.}
    \label{fig:gauspcc-1k}
    \vspace{-10pt}
\end{figure}

\textbf{Local Density Analysis.} Local neighborhoods are crucial for achieving efficient compression in the model. Excessively dense or sparse neighborhoods can hinder the point cloud feature extractors from effectively capturing local features, thereby affecting the model's ability to build contextual information. We begin by conducting local density analysis on existing point cloud datasets, including 8iVFB \cite{dataset8i}, Kitti \cite{kitti}, ScanNet \cite{dai2017scannet}, and 3DGS point clouds \cite{barron2022mipnerf360,knapitsch2017tanks,hedman2018deep}. We calculate the number of neighbors within a $k$-neighborhood at different scales to measure the local density of point clouds, with detailed calculation processes provided in Appendix~\ref{Local Density Analysis}. We visualize the neighborhood density across different datasets in Figure \ref{fig:pc_vis}, where increasing redness indicates regions of higher point concentration, and present the local density distributions of various datasets in Figure \ref{fig:pc_ld_fd} and Appendix Figure \ref{fig:all_scale_local_density}. As shown, existing point clouds at certain scales (8iVFB, Kitti, ScanNet) exhibit a distribution characterized by higher density in the middle and lower at both ends, indicating that their local densities are concentrated around the mean value. In contrast, Gaussian point clouds demonstrate a more gradual distribution, suggesting that neighbor counts in Gaussian point clouds span across various quantities, with both sparse and dense neighborhoods coexisting locally. This characteristic distinguishes the geometric distribution of Gaussian point clouds from that of traditional point clouds, making it more difficult for local feature extractors to extract features from Gaussian point clouds and increasing the challenge of building contextual information.

\textbf{Fractal Dimension Analysis.} Fractal dimension characterizes the global density distribution of point clouds and influences the context construction of entropy models in the mainstream hierarchical prediction-based methods in the current point cloud compression field.
As shown in Figure \ref{fig:pc_ld_fd}, we present the fractal dimensions of all datasets, with detailed calculation processes provided in Appendix~\ref{Fractal Dimension Analysis}. The graph reveals that the fractal dimensions of Gaussian point clouds closely resemble those of Kitti after Scale 5, suggesting that Gaussian point clouds share similar globally sparse density characteristics with Kitti. Therefore, we can conclude that Gaussian point clouds exhibit locally dense yet globally sparse density characteristics, significantly different from conventional point clouds. This difference results in suboptimal compression efficiency when using traditional compression algorithms.
Furthermore, considering the principles of 3DGS Gaussian representation \cite{3dgs,lu2024scaffold}, scene centers require dense Gaussian primitives for modeling, while background areas can be modeled with sparse Gaussian primitives. This theoretical understanding aligns perfectly with the conclusions drawn from our data analysis.

\textbf{Mismatch between Current and Gaussian Datasets.} The statistical distribution $P_{\text{Gaussian}}(\mathbf{x})$ of Gaussian point clouds differs substantially from conventional point clouds $P_{\text{Conv}}(\mathbf{x})$, creating a significant domain gap. The subsequent experimental results, as shown in Table \ref{table:pcc_res}, also demonstrate that models trained on Kitti perform poorly when directly applied to Gaussian point clouds. Therefore, it is necessary to propose a Gaussian point cloud dataset for training AI compression models.

\textbf{Data Collection and Generation.} We examine challenges in applying AI-based point cloud compression to Gaussian point clouds, primarily stemming from distributional differences between conventional and Gaussian point cloud datasets.
While AI methods excel with datasets like 8iVFB and Kitti due to similar large-scale training data, Gaussian point clouds lack comparable datasets, limiting AI performance in this domain. To address this gap, we introduce GausPcc-1K, a comprehensive collection of pre-trained 3D scenes (Figure \ref{fig:gauspcc-1k}), enabling effective application of point cloud compression techniques to Gaussian data.
We employ \textbf{Scaffold-GS} \cite{lu2024scaffold} for 3D scene training due to its state-of-the-art performance in structured 3D representation. As illustrated in Figure \ref{fig:gauspcc-1k}, this framework enables us to construct GausPcc-1K by selecting a representative subset from the DL3DV \cite{ling2024dl3dv} dataset, from which we meticulously curate 1,000 high-fidelity pre-trained 3DGS models exhibiting an average PSNR of 29.01dB. The dataset creation process is executed on a single RTX 3090 GPU, consuming approximately 40 days of computational resources. We systematically train in excess of 1,000 samples and employ rigorous quality assessment criteria to select the optimal 1,000 specimens that constitute the final collection. Subsequently, we extract positions and feature vectors from these models to establish the comprehensive GausPcc-1K Gaussian point cloud dataset.

\begin{figure}
    \centering
    \includegraphics[width=1.0\linewidth]{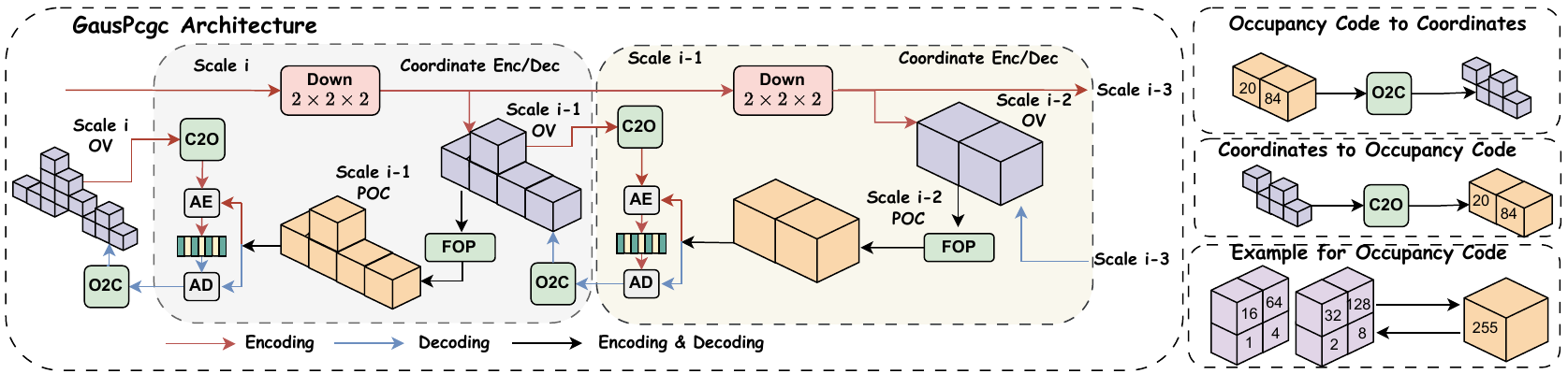}
    \caption{llustration of the proposed GausPcgc framework.}
    \label{fig:gauspcc}
    \vspace{-8pt}
\end{figure}

\textbf{Dataset Characteristics Analysis.} Here we introduce the characteristics of the proposed GausPcc-1K dataset. This collection is distinguished by its high fidelity, comprehensive distribution, significant diversity, and novel geometric distribution characteristics. Moreover, it exhibits congruence with the intrinsic geometric properties of Gaussian point clouds, thereby providing appropriate training data for AI-based point cloud geometry compression methodologies.
As demonstrated in Figure \ref{fig:pc_ld_fd}, GausPcc-1K shares distribution characteristics with commonly used Gaussian point cloud test sets, specifically the locally dense yet globally sparse geometric arrangement. This alignment makes our dataset particularly suitable for Gaussian point cloud compression research. Additionally, GausPcc-1K preserves attribute values beyond geometry, establishing a foundation for the development of subsequent attribute compression methods.

\subsection{GausPcgc Framework}
\label{sub: GausPcc Method}

\textbf{Hierarchical Occupancy Code Prediction.}
Inspired by RENO \cite{you2025reno}, we adopt occupancy codes for efficient point cloud representation. Occupancy codes provide an efficient spatial representation in 3D space. When using 8-bit occupancy codes, each voxel can represent the occupancy status of its eight child voxels. Specifically, an occupancy code at scale $i$ encapsulates the voxel occupancy information at scale $i+1$, as illustrated in Figure \ref{fig:gauspcc}. We denote occupied voxels as $\mathrm{OV}$, occupancy codes as $\mathrm{OC}$, and the predicted probability distribution of occupancy codes as $\mathrm{POC}$.

Our compression framework operates in multiple stages. First, we transform the point cloud into a multi-scale voxelized representation. Then, we predict the probability distribution of occupancy codes at each scale in a progressive manner. Finally, these probabilities are used for lossless entropy coding through arithmetic coding. The entire process is depicted in Figure \ref{fig:gauspcc}. Thus, our problem is reformulated as accurately predicting occupancy code probabilities—the more accurate the probability estimation, the smaller the resulting bitstream.

\textbf{Four-stage Occupancy Predictor (FOP).}
\begin{table}[]
\caption{Comparison of compression efficiency of different compression algorithms, with the best and second-best results highlighted in \colorbox[HTML]{FFC7CE}{\textcolor{black}{red}} and \colorbox[HTML]{FFEB9C}{\textcolor{black}{yellow}} cells.}
\label{table:pcc_res}
\centering
\resizebox{0.9\textwidth}{!}{
\setlength{\tabcolsep}{3pt} 
\begin{tabular}{@{}c|c|cc|cc|cc|cc|c@{}}
\toprule
\textbf{Model}   & \textbf{G-PCC v23}                                   & \multicolumn{2}{c|}{\textbf{SparsePCGC}} & \multicolumn{2}{c|}{\textbf{Octattention}}                     & \multicolumn{2}{c|}{\textbf{EHEM}} & \multicolumn{2}{c|}{\textbf{RENO}} & \textbf{Ours} \\ \hline
Test/Train       & /                                                    & Kitti               & Ours            & Kitti   & Ours                                              & Kitti            & Ours         & Kitti            & Ours         & Ours       \\ \toprule
bicycle          & 15.69                                                & 20.74               & 18.81              & 18.72   & 10.59                                                & 19.39            & 12.83           & 19.55            & 13.40           & 12.31         \\
bonsai           & 17.20                                                & 18.77               & 17.42              & 18.17   & 14.87                                                & 17.93            & 15.98           & 18.13            & 16.07           & 15.33         \\
counter          & 15.61                                                & 17.07               & 15.92              & 16.61   & 13.79                                                & 16.37            & 15.07           & 16.62            & 16.08           & 15.40         \\
drjohnson        & 9.96                                                 & 11.54               & 10.42              & 10.95   & 8.94                                                 & 11.02            & 9.55            & 11.25            & 10.32           & 9.64          \\
flowers          & 15.73                                                & 21.04               & 19.29              & 18.98   & 10.42                                                & 19.75            & 13.50           & 19.89            & 13.44           & 12.38         \\
garden           & 15.75                                                & 19.43               & 17.98              & 18.26   & 11.61                                                & 18.63            & 13.64           & 18.65            & 13.72           & 12.78         \\
kitchen          & 14.16                                                & 15.84               & 13.92              & 14.61   & 12.81                                                & 14.53            & 13.56           & 14.81            & 14.32           & 13.65         \\
playroom         & 11.74                                                & 14.12               & 12.80              & 13.03   & 10.11                                                & 13.26            & 12.48           & 13.60            & 13.71           & 13.37         \\
room             & 17.13                                                & 19.11               & 17.61              & 18.39   & 14.77                                                & 18.15            & 16.69           & 18.48            & 17.56           & 16.66         \\
stump            & 18.27                                                & 22.66               & 20.79              & 20.96   & 12.41                                                & 21.11            & 17.63           & 21.65            & 17.73           & 17.00         \\
train            & 9.58                                                 & 10.64               & 9.32               & 9.57    & 7.62                                                 & 9.76             & 8.55            & 9.79             & 8.23            & 7.76          \\
treehill         & 15.54                                                & 21.42               & 19.13              & 18.86   & {10.50}                                                & 19.71            & 12.62           & 19.82            & 13.48           & 12.38         \\
truck            & 11.63                                                & 13.40               & 12.29              & 12.25   & {8.62}                                                 & 12.69            & 11.76           & 12.94            & 12.53           & 12.53         \\ \hline
Avg Bpp $\downarrow$             & 14.46 & 17.37               & 15.82              & 16.10   &  \cellcolor[HTML]{FFC7CE}{11.31} & 16.33            & 13.37           & 16.55            & 13.89           & \cellcolor[HTML]{FFEB9C}{13.27}         \\ \hline
\textbf{CR Gain} $\downarrow$ & 0.0\%                                              & 20.1\%             & 9.4\%             & 11.4\% & -21.8\%                                             & 12.9\%          & -7.5\%         & 14.50\%          & -3.9\%         & -8.2\%       \\ \hline
\textbf{Enc/Dec} $\downarrow$  & {5.72/4.04}                                   & \multicolumn{2}{c|}{{3.81/3.97}} & \multicolumn{2}{c|}{{19.9/3414.6}}                     & \multicolumn{2}{c|}{{12.6/12.5}} & \multicolumn{2}{c|}{\cellcolor[HTML]{FFC7CE}{0.517/0.551}} & \cellcolor[HTML]{FFEB9C}{0.797/0.834} \\ 
\toprule
\end{tabular}
}
\vspace{-6pt}
\end{table}
To enhance occupancy code probability prediction accuracy, we propose a Four-stage Occupancy Predictor (FOP), as shown in  Appendix Figure \ref{fig:fop}. The prediction process leverages both inter-scale and intra-scale contexts:
\begin{equation}
P(\mathrm{OC}_i | \mathcal{C}_i) = f_{\theta}(\mathrm{OC}_{i-1}, \mathcal{C}_i),
\end{equation}
 where $\mathrm{OC}_i \in \{0, 1, \ldots, 255\}$ represents the occupancy code at scale $i$, $\mathcal{C}_i$ denotes the corresponding coordinates, and $f_{\theta}$ is our prediction network with parameters $\theta$.
For 8-bit occupancy code prediction, we introduce a non-uniform grouping schemes \cite{unipcgc} that divides the prediction task into multiple stages. Instead of directly predicting all 8 bits simultaneously (which would require modeling $2^8=256$ possible states), we decompose the prediction into stages with conditional dependencies:
\begin{equation}
P(\mathrm{OC}_i | \mathcal{C}_i) = P(b_1 | \mathcal{C}_i) \cdot P(b_2 | b_1, \mathcal{C}_i) \cdot P(b_{3:4} | b_1, b_2, \mathcal{C}_i) \cdot P(b_{5:8} | b_1, b_2, b_{3:4}, \mathcal{C}_i),
\end{equation}
 where $b_j$ represents the $j$-th bit of the occupancy code, and $b_{a:b}$ denotes bits from position $a$ to $b$ inclusive.
The prediction network architecture incorporates spatial convolutions at each stage to capture local neighbor prior information. Let $\mathbf{F}_i$ represent the feature tensor at scale $i$, and $\mathbf{b}_{<j}$ denote the previously predicted bits before stage $j$. The prediction process can be formulated recursively for each stage $j \in \{1,2,3,4\}$ as:
\begin{equation}
\begin{aligned}
\mathbf{F}_{i,j} = \mathcal{S}_j(\mathbf{F}_i + \mathcal{E}_j(\mathbf{b}_{<j})), \\
P(\mathbf{b}_j | \mathbf{b}_{<j}, \mathcal{C}_i) = \sigma(\mathcal{H}_j(\mathbf{F}_{i,j})),
\end{aligned}
\end{equation}
 where $\mathcal{S}_j$ represents the spatial convolution network, $\mathcal{H}_j$ is the prediction head, $\mathcal{E}_j$ is the bit embedding function (with $\mathcal{E}_1 = 0$ for the first stage), and $\sigma$ denotes the softmax activation. Each stage progressively predicts different bit groups: $\mathbf{b}_1 = b_1$, $\mathbf{b}_2 = b_2$, $\mathbf{b}_3 = b_{3:4}$, and $\mathbf{b}_4 = b_{5:8}$, with increasing context information.
The information-theoretic entropy of each prediction is calculated as:
\begin{equation}
\mathcal{L}_{\text{entropy}}(j, v) = -\log_2(P(b_j^v | b_{<j}^v, \mathcal{C}_i) + \epsilon),
\end{equation}
 where $b_j^v$ represents the bits predicted at stage $j$ for voxel $v$, $b_{<j}^v$ denotes all previously predicted bits for the same voxel, and $\epsilon$ is a small constant (typically $10^{-10}$) added for numerical stability.
The total bitrate for the point cloud is the sum of entropies across all scales, stages, and voxels, divided by the total number of points $N$:
\begin{equation}
\text{Bpp} = \frac{1}{N}\sum_{i=1}^{M}\sum_{j=1}^{4}\sum_{v \in \mathcal{V}_i}\mathcal{L}_{\text{entropy}}(j, v),
\end{equation} where $M$ is the number of scales, and $\mathcal{V}_i$ is the set of voxels at scale $i$.
This progressive, conditional prediction approach significantly improves coding efficiency by exploiting both spatial correlations through 3D convolutions and bit-level statistical dependencies through the conditional probability model. Additional details are provided in the Appendix \ref{app:gauspcc_method}.

\section{Experiments}
\label{sec:exper}
\begin{table}[]
\caption{Performance comparison across different datasets and methods, with the best and second-best results highlighted in \colorbox[HTML]{FFC7CE}{\textcolor{black}{red}} and \colorbox[HTML]{FFEB9C}{\textcolor{black}{yellow}} cells.}
\label{table:gszip}
\centering
\resizebox{1.0\textwidth}{!}{
\setlength{\tabcolsep}{3pt} 
\begin{tabular}{c|cccc|cccc|cccc}
\toprule
                                                                                     & \multicolumn{4}{c|}{\textbf{Mip-NeRF360}} & \multicolumn{4}{c|}{\textbf{Tanks and Temples}} & \multicolumn{4}{c}{\textbf{Deep Blending}} \\
{\textbf{\begin{tabular}[c]{@{}c@{}}Datasets\\ Models\end{tabular}}} & PSNR$\uparrow$     & SSIM$\uparrow$     & LPIPS$\downarrow$    & Size$\downarrow$     & PSNR$\uparrow$       & SSIM$\uparrow$       & LPIPS$\downarrow$     & Size$\downarrow$      & PSNR$\uparrow$      & SSIM$\uparrow$     & LPIPS$\downarrow$    & Size$\downarrow$     \\ \toprule
\textbf{3DGS} \cite{3dgs}                                                                        & 27.46    & \cellcolor[HTML]{FFC7CE}{0.812}    & 0.222    & 750.9    & 23.69      & 0.844      & 0.178     & 431       & 29.42     & 0.899    & 0.247    & 663.9    \\
\textbf{Scaffold-GS} \cite{lu2024scaffold}                                                                 & 27.5     & 0.806    & 0.252    & 253.9    & 23.96      & 0.853      & 0.177     & 86.5      & 30.21     & 0.906    & 0.254    & 66       \\ \hline
\textbf{Compact3D} \cite{lee2024compact}                                                                  & 27.08    & 0.798    & 0.247    & 48.8     & 23.32      & 0.831      & 0.201     & 39.43     & 29.79     & 0.901    & 0.258    & 43.21    \\
\textbf{SOG} \cite{morgenstern2024compact}                                                                         & 26.56    & 0.791    & 0.241    & 16.7     & 23.15      & 0.828      & 0.198     & 9.3       & 29.12     & 0.892    & 0.270     & 5.7      \\
\textbf{Compressed3D} \cite{niedermayr2024compressed}                                                                & 26.98    & 0.801    & 0.238    & 28.8     & 23.32      & 0.832      & 0.194     & 17.28     & 29.38     & 0.898    & 0.253    & 25.3     \\
\textbf{RDOGaussian} \cite{wang2024end}                                                                 & 27.05    & 0.802    & 0.239    & 23.5     & 23.34      & 0.835      & 0.195     & 12.03     & 29.63     & 0.902    & 0.252    & 18       \\
\textbf{ContextGS} \cite{wang2024contextgs}                                                                   & 27.62    & 0.808    & 0.237    & 12.7     & 24.20       & 0.852      & 0.184     & 7.05      & 30.11     & 0.907    & 0.265    & 3.45     \\ 
\textbf{CompGS} \cite{liu2024compgs} & 26.37    & 0.778    & 0.276    & 8.83     & 23.11       & 0.815      & 0.236     & 5.89      & 29.30     & 0.895    & 0.293    & 6.03  \\
\textbf{Reduced3DGS} \cite{papantonakis2024reducing} & 27.19    & 0.807    & 0.230    & 29.54 & 23.57       & 0.840      & 0.188     & 14.00      & 29.63     & 0.902    & 0.249    & 18.00  \\
\textbf{LightGaussian} \cite{fan2024lightgaussian} & 27.00    & 0.799    & 0.249    & 44.54 & 22.83       & 0.822      & 0.242     & 22.43      & 27.01     & 0.872    & 0.308    & 33.94  \\
\hline
\textbf{TC-GS} \cite{wang2025tc} & 27.61    & 0.801    & \cellcolor[HTML]{FFC7CE}{0.166}    & 13.85    & 23.94      & 0.843      & \cellcolor[HTML]{FFC7CE}{0.113}     & 7.89      & 30.04     & 0.899    & \cellcolor[HTML]{FFEB9C}{0.122}    & 4.20     \\
\textbf{HAC (low)} \cite{chen2024hac}                                                                   & 27.55    & 0.807    & 0.239    & 15.23    & 24.29      & 0.850      & 0.185     & 8.06      & 30.06     & 0.907    & 0.267    & 4.31     \\
\textbf{HAC (high)} \cite{chen2024hac}                                                                  & 27.82    & \cellcolor[HTML]{FFEB9C}{0.811}    & 0.229    & 25.27    & 24.36      & \cellcolor[HTML]{FFC7CE}{0.857}      & 0.174     & 13.22     & 30.27     & \cellcolor[HTML]{FFEB9C}{0.910}    & 0.255    & 7.65     \\
\textbf{HAC++ (low)} \cite{chen2025hac++}                                                                 & 27.54    & 0.802    & 0.253    & 8.37     & 24.30       & 0.850       & 0.189     & 5.17      & 30.13     & 0.907    & 0.265    & 2.89     \\
\textbf{HAC++ (high)} \cite{chen2025hac++}                                                               & \cellcolor[HTML]{FFEB9C}{27.80}     & \cellcolor[HTML]{FFEB9C}{0.811}    & 0.230     & 18.57    & 24.28      & \cellcolor[HTML]{FFEB9C}{0.856}      & 0.173     & 10.52     & \cellcolor[HTML]{FFC7CE}{30.34}     & \cellcolor[HTML]{FFEB9C}{0.910}     & 0.253    & 6.71      \\
\textbf{CAT-3DGS (low)} \cite{zhancat}                                                              & 27.14    & 0.791    & 0.279    & \cellcolor[HTML]{FFEB9C}{5.51}     & 24.20      & 0.838      & 0.217     & \cellcolor[HTML]{FFEB9C}{3.57}      & 29.64     & 0.900     & 0.294    & \cellcolor[HTML]{FFEB9C}{1.95}     \\
\textbf{CAT-3DGS (high)} \cite{zhancat}                                                            & 27.70    & 0.808    & 0.246    & 12.36    & 24.38      & 0.850      & 0.195     & 6.73      & 30.18     & \cellcolor[HTML]{FFEB9C}{0.910}     & 0.273    & 3.65     \\ \hline
\textbf{Ours-TC-GS} & 27.59    & 0.800    & \cellcolor[HTML]{FFEB9C}{0.167}    & 11.94    & 23.88      & 0.838      & \cellcolor[HTML]{FFEB9C}{0.117}     & 6.74      & 30.14     & 0.901    & \cellcolor[HTML]{FFC7CE}{0.119}    & 3.57     \\
\cellcolor[HTML]{FFFFFF}\textbf{Ours-HAC (low)}                                   & 27.54    & 0.807    & 0.239    & 12.48    & 24.23      & 0.849      & 0.187     & 6.53      & 30.11     & 0.906    & 0.266    & 3.56     \\
\cellcolor[HTML]{FFFFFF}\textbf{Ours-HAC (high)}                                  & \cellcolor[HTML]{FFC7CE}{27.86}    & \cellcolor[HTML]{FFEB9C}{0.811}    & 0.229    & 22.03    & \cellcolor[HTML]{FFC7CE}{24.52}      & \cellcolor[HTML]{FFC7CE}{0.857}      & 0.175     & 10.87     & 30.30     & \cellcolor[HTML]{FFC7CE}{0.911}    & 0.254    & 6.63     \\
\cellcolor[HTML]{FFFFFF}\textbf{Ours-HAC++ (low)}                                 & 27.58    & 0.803    & 0.252    & 8.18     & 24.18      & 0.848      & 0.189     & 5.22      & 30.17     & 0.907    & 0.266    & 2.83     \\
\cellcolor[HTML]{FFFFFF}\textbf{Ours-HAC++ (high)}                                & \cellcolor[HTML]{FFEB9C}{27.80}     & \cellcolor[HTML]{FFEB9C}{0.811}    & 0.231    & 18.25    & 24.33      & \cellcolor[HTML]{FFEB9C}{0.856}      & 0.174     & 10.3      & \cellcolor[HTML]{FFEB9C}{30.31}     & \cellcolor[HTML]{FFEB9C}{0.910}     & 0.254    & 6.69     \\
\cellcolor[HTML]{FFFFFF}\textbf{Ours-CAT (low)}                              & 27.13    & 0.790    & 0.281    & \cellcolor[HTML]{FFC7CE}{4.33}     & 24.12      & 0.836      & 0.219     & \cellcolor[HTML]{FFC7CE}{2.87}      & 29.81     & 0.900     & 0.294    & \cellcolor[HTML]{FFC7CE}{1.56}     \\
\cellcolor[HTML]{FFFFFF}\textbf{Ours-CAT (high)}                             & 27.70    & 0.808    & 0.247    & 10.42    & \cellcolor[HTML]{FFEB9C}{24.45}      & 0.850      & 0.195     & 5.49      & 30.10     & \cellcolor[HTML]{FFEB9C}{0.910}     & 0.273    & 3.12     \\ \toprule
\end{tabular}
}
\vspace{-6pt}
\end{table}
\subsection{Experiment Setup}
\label{sub:Experiment Setup}
\textbf{Dataset.} We evaluate our method using three commonly used 3DGS benchmark datasets: Mip-NeRF360 \cite{barron2022mipnerf360}, Tanks and Temples \cite{knapitsch2017tanks}, and Deep Blending \cite{hedman2018deep}. We extract geometric positions from trained 3DGS models as point cloud compression evaluation datasets. For Gaussian compression evaluation, we align with current popular evaluation protocols. Our training datasets include Kitti \cite{kitti} and our custom GausPcc-1K dataset. In our paper, ``Gaussian'' in legends indicates point clouds from standard 3DGS test sets, while ``GausPcc'' represents our proposed dataset.

\textbf{Baseline.} For point cloud compression baselines, we compare against the latest traditional compression method G-PCC v23 \cite{gpcc} and AI-based methods including SparsePCGC \cite{sparsepcgc}, Octattention \cite{fu2022octattention}, EHEM \cite{song2023efficient}, and RENO \cite{you2025reno}. For Gaussian compression, we benchmark against popular methods such as 3DGS \cite{3dgs}, Scaffold-GS \cite{lu2024scaffold}, Compact3D \cite{lee2024compact}, SOG \cite{morgenstern2024compact}, Compressed3D \cite{niedermayr2024compressed}, RDOGaussian \cite{wang2024end}, ContextGS \cite{wang2024contextgs}, CompGS \cite{liu2024compgs}, Reduced3DGS \cite{papantonakis2024reducing}, LightGaussian \cite{fan2024lightgaussian}, HAC \cite{chen2024hac}, HAC++ \cite{chen2025hac++}, Cat-3DGS \cite{zhancat} and TC-GS \cite{wang2025tc}.

\textbf{Implementation.} All experiments are conducted on a single NVIDIA RTX 3090 GPU and an Intel(R) Xeon(R) Silver 4314 CPU. The creation of the GausPcc-1K dataset requires approximately 40 days of GPU computation time. For point cloud compression, we implement our approach using the torchsparse \cite{tangandyang2023torchsparse} and train the model for $10^6$ iterations, consuming 48 hours of GPU time. Our Gaussian compression method maintains consistency with the original training protocol. Detailed training procedures and hyperparameters are provided in the supplementary materials.

\subsection{Benchmark Results}
\textbf{Point Cloud Compression.}
We compare leading open-source point cloud compression methods, including EHEM (unofficial implementation \cite{luo2024scp}). Our analysis shows Kitti and Gaussian point clouds have similar geometric distributions, so we trained these methods on both Kitti and our GausPcc-1K dataset.
Models trained on Kitti fail to outperform G-PCC, but when retrained on GausPcc-1K, most methods surpass G-PCC, validating our dataset's effectiveness. Octattention \cite{fu2022octattention} achieves best performance on GausPcc-1K, while RENO \cite{you2025reno} offers fastest inference. Our approach provides the second-best performance with near-optimal speed, improving 8.2\% over G-PCC v23 with 6× faster inference.
Table \ref{table:pcc_res} establishes a new benchmark for Gaussian point cloud compression, defining this specialized task and demonstrating point cloud compression can be optimized for Gaussian geometry. We conduct a more comprehensive analysis of the point cloud compression experimental results, examining performance across Gaussian point clouds, Kitti point clouds, and 8iVFB point clouds, as detailed in Appendix \ref{app:sub:current_ana} and \ref{app:sub: gaussian_ana}.
\begin{table}[]
\caption{Detailed bitstream analysis for the Mip-NeRF 360 \cite{barron2022mipnerf360} bicycle scene. }
\label{table:size_hac}
\centering
\resizebox{0.85\textwidth}{!}{
\setlength{\tabcolsep}{3pt} 

\begin{tabular}{c|c|cccccccc|cc}
\toprule
\multirow{2}{*}{\textbf{}} & \multirow{2}{*}{\begin{tabular}[c]{@{}c@{}}Number of \\ Anchors\end{tabular}} & \multicolumn{8}{c|}{Storage Cost   (MB)}                                  & \multicolumn{2}{c}{Fidelity} \\ \cline{3-12} 
                           &                                                                               & Position & Feat    & Scaling & Offsets & Hash   & Masks  & MLPs   & Total   & PSNR          & SSIM         \\ \toprule
HAC                        & 912838                                                                        & 5.22 & 22.77  & 4.94  & 11.34 & 0.13 & 1.09 & 0.16 & 45.67 & 25.01         & 0.743        \\
Ours-HAC                   & 819267                                                                        & 1.08 & 19.69 & 4.43  & 9.91  & 0.14  & 0.98 & 0.16 & 36.39 & 25.14         & 0.744        \\ \toprule
\end{tabular}
}
\vspace{-10pt}
\end{table}

\textbf{Remark.} Despite Octattention showing superior compression efficiency, its decoding process requires 3414.6 seconds due to its autoregressive inference approach. This prohibitive decoding time renders the method impractical for real-world applications. More analysis are shown in Appendix \ref{app:sub: gaussian_ana}.

\textbf{Gaussian Compression.}
Integrating our proposed GausPcgc into HAC \cite{chen2024hac}, HAC++ \cite{chen2025hac++}, Cat-3DGS \cite{zhancat} and TC-GS \cite{wang2025tc} enhances their performance without increasing inference latency, as shown in Table \ref{table:gszip}. We also present rate-distortion curves in Appendix Figure \ref{fig:rd_res}. For example, GausPcc-HAC achieves a 0.06dB PSNR improvement while reducing model size by 3.24MB, representing a 12.8\% size reduction compared to the original model. This performance gain primarily results from the efficient compression of geometric positions. HAC++ shows limited performance improvement since it already employs G-PCC as a geometric position compressor; however, using GausPcgc for compression still reduces bitstream size and further accelerates inference speed. As an enhancer, GausPcgc provides additional compression for 3DGS without introducing encoding or decoding delays. We further present the bitstream size by component in Table \ref{table:size_hac}. For more detailed analysis, please refer to Appendix \ref{app:sub:bits ana}.

\begin{figure}[htbp]
    \centering
    \begin{subfigure}[b]{0.19\textwidth}
        \centering
        \includegraphics[width=\textwidth]{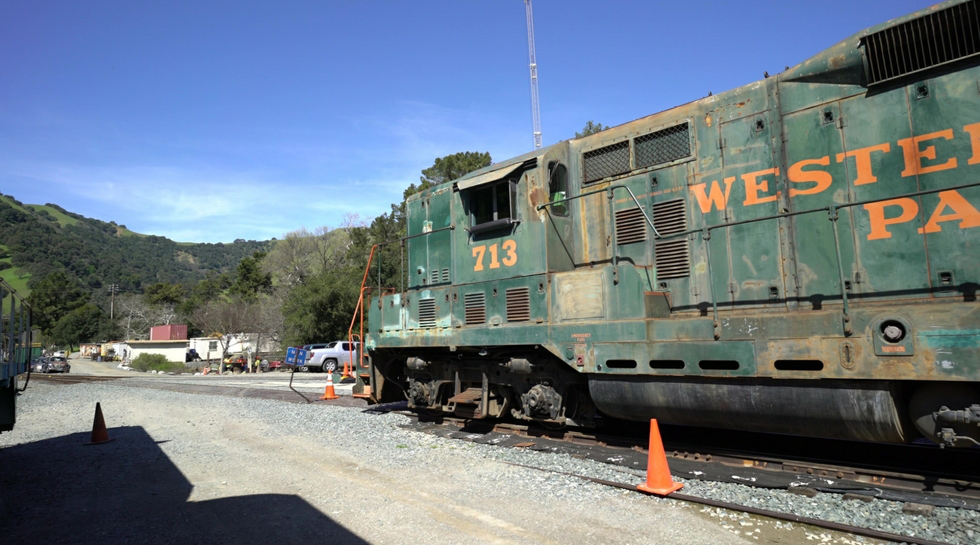}
        \caption{Ground \\ Truth}
        \label{fig:gt-00004}
    \end{subfigure}
    \hfill
    \begin{subfigure}[b]{0.19\textwidth}
        \centering
        \includegraphics[width=\textwidth]{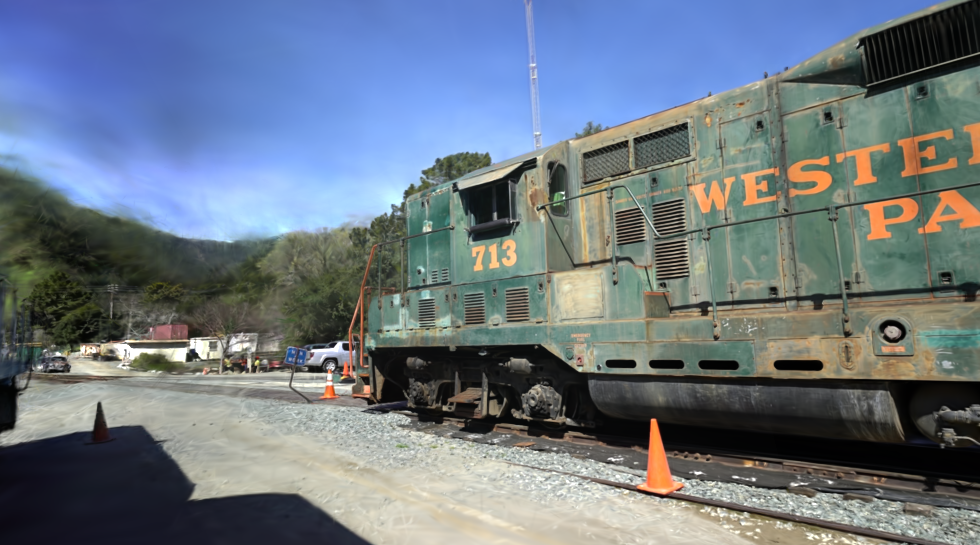}
        \caption{3DGS\\22.03dB/257.3MB}
        \label{fig:3dgs-00004}
    \end{subfigure}
    \hfill
    \begin{subfigure}[b]{0.19\textwidth}
        \centering
        \includegraphics[width=\textwidth]{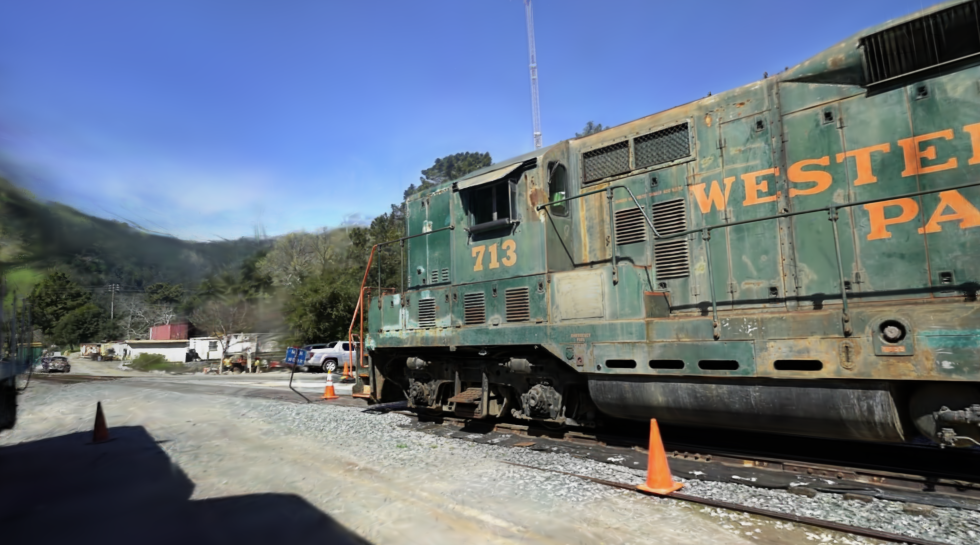}
        \caption{Scaffold-GS\\22.37dB/92.24MB}
        \label{fig:scaf-00004}
    \end{subfigure}
    \hfill
    \begin{subfigure}[b]{0.19\textwidth}
        \centering
        \includegraphics[width=\textwidth]{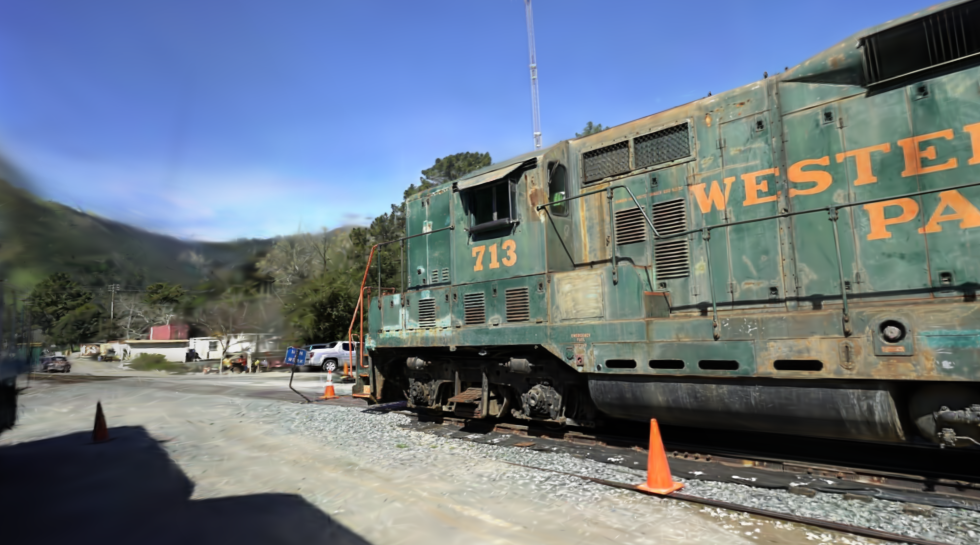}
        \caption{Ours-TC-GS\\22.15dB/6.22MB}
        \label{fig:our-tcgs-00004}
    \end{subfigure}
    \hfill
    \begin{subfigure}[b]{0.19\textwidth}
        \centering
        \includegraphics[width=\textwidth]{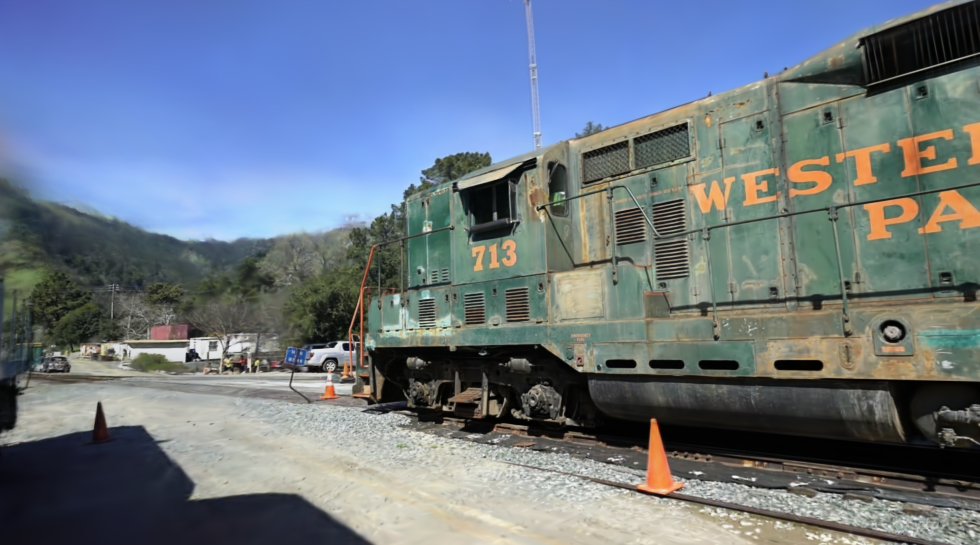}
        \caption{Ours-HAC\\22.94dB/9.17MB}
        \label{fig:our-hac-00004}
    \end{subfigure}
    \caption{Visual comparison on Tanks and Temples \cite{knapitsch2017tanks} train scene.}
    \label{fig:main_tex_train_vis}
    \vspace{-8pt}
\end{figure}

\textbf{Visualization.} We present qualitative comparisons of visual quality in the Figure \ref{fig:main_tex_train_vis}, Appendix Figure \ref{fig:bicycle_vis} and Figure \ref{fig:train_vis}, demonstrating that our proposed method achieves superior perceptual quality compared to the original 3DGS \cite{3dgs} while achieving a compression ratio of over 30×.

\subsection{Ablation Experiments}
\textbf{Dataset Effectiveness.} We train current AI-based point cloud compression methods using both Kitti and GausPcc-1K datasets separately. The results demonstrate significant performance improvements across all methods, with Octattention achieving a remarkable 30\% gain, as shown in Table \ref{table:pcc_res}. These improvements clearly validate the effectiveness of our proposed GausPcc-1K dataset.

\textbf{Module Effectiveness.} 
Table \ref{table:abla} presents our ablation study results, where UE refers to the non-uniform grouping strategy and NP indicates the utilization of neighbor prior information. We found that incorporating neighbor priors enhances compression efficiency with minimal impact on inference latency. Furthermore, using 4-stage grouping demonstrates superior encoding performance compared to 2-stage grouping. Notably, replacing standard 4-stage grouping with our non-uniform grouping strategy further improves encoding performance without affecting codec speed. These findings clearly demonstrate the effectiveness of the modules proposed in this paper.
\begin{table}[]
\centering
\caption{Ablation experiment of the modules used by GausPcgc.}
\label{table:abla}
\resizebox{0.75\textwidth}{!}{
\begin{tabular}{c|cccc|ccc}
\toprule
\textbf{}            & 2stage & 4stage & UE & NP & Bpp    & Enc times & Dec times \\ \toprule
\textbf{RENO}        & \checkmark      &  \texttimes      &  \texttimes  & \texttimes   & 13.890  & 0.517     & 0.551     \\
\textbf{RENO+NP}     & \checkmark      &   \texttimes     &  \texttimes  & \checkmark  & 13.458 & 0.604     & 0.637     \\
\textbf{RENO-4stage} & \checkmark      & \checkmark      &  \texttimes  & \texttimes   & 13.609 & 0.547     & 0.586     \\
\textbf{Ours w/o UE} &   \texttimes     & \checkmark      &  \texttimes  & \checkmark  & 13.379 & 0.799     & 0.837      \\
\textbf{Ours}        &   \texttimes     & \checkmark      & \checkmark & \checkmark  & 13.274 & 0.797     & 0.834     \\ \toprule
\end{tabular}
}
\vspace{-6pt}
\end{table}
\section{Conclusion}
This paper presents the first benchmark, dataset, and method for Gaussian point cloud geometry compression, providing an efficient geometric positions compression scheme for current Gaussian compression approaches. We investigate the reasons behind the inefficiency of existing AI-based point cloud compression methods and introduce the novel \textbf{\textit{GausPcc-1K}} dataset and \textbf{\textit{GausPcgc}} method specifically designed for Gaussian point cloud compression. Our proposed method further enhances the efficiency of existing Gaussian compression without compromising coding speeds.

{
\small
\bibliographystyle{abbrv}
\bibliography{Reference.bib}

\begin{thebibliography}{10}

\bibitem{3dgszip}
M.~T. Bagdasarian, P.~Knoll, Y.-H. Li, F.~Barthel, A.~Hilsmann, P.~Eisert, and W.~Morgenstern.
\newblock 3dgs. zip: A survey on 3d gaussian splatting compression methods.
\newblock {\em arXiv preprint arXiv:2407.09510}, 2024.

\bibitem{barron2022mipnerf360}
J.~T. Barron, B.~Mildenhall, D.~Verbin, P.~P. Srinivasan, and P.~Hedman.
\newblock Mip-nerf 360: Unbounded anti-aliased neural radiance fields.
\newblock In {\em Proceedings of the IEEE/CVF conference on computer vision and pattern recognition}, pages 5470--5479, 2022.

\bibitem{gpcc}
C.~Cao, M.~Preda, V.~Zakharchenko, E.~S. Jang, and T.~Zaharia.
\newblock Compression of sparse and dense dynamic point clouds—methods and standards.
\newblock {\em Proceedings of the IEEE}, 109(9):1537--1558, 2021.

\bibitem{chen2022tensorf}
A.~Chen, Z.~Xu, A.~Geiger, J.~Yu, and H.~Su.
\newblock Tensorf: Tensorial radiance fields.
\newblock In {\em European conference on computer vision}, pages 333--350. Springer, 2022.

\bibitem{chen2025pcgs}
Y.~Chen, M.~Li, Q.~Wu, W.~Lin, M.~Harandi, and J.~Cai.
\newblock Pcgs: Progressive compression of 3d gaussian splatting.
\newblock {\em arXiv preprint arXiv:2503.08511}, 2025.

\bibitem{chen2024hac}
Y.~Chen, Q.~Wu, W.~Lin, M.~Harandi, and J.~Cai.
\newblock Hac: Hash-grid assisted context for 3d gaussian splatting compression.
\newblock In {\em European Conference on Computer Vision}, pages 422--438. Springer, 2024.

\bibitem{chen2025hac++}
Y.~Chen, Q.~Wu, W.~Lin, M.~Harandi, and J.~Cai.
\newblock Hac++: Towards 100x compression of 3d gaussian splatting.
\newblock {\em arXiv preprint arXiv:2501.12255}, 2025.

\bibitem{dai2017scannet}
A.~Dai, A.~X. Chang, M.~Savva, M.~Halber, T.~Funkhouser, and M.~Nie{\ss}ner.
\newblock Scannet: Richly-annotated 3d reconstructions of indoor scenes.
\newblock In {\em Proceedings of the IEEE conference on computer vision and pattern recognition}, pages 5828--5839, 2017.

\bibitem{dataset8i}
E.~d'Eon, H.~Bob, T.~Myers, and P.~A. Chou.
\newblock 8i voxelized full bodies - a voxelized point cloud dataset.
\newblock In {\em ISO/IEC JTC1/SC29 Joint WG11/WG1 (MPEG/JPEG) input document WG11M40059/WG1M74006}, 2017.

\bibitem{fan2024lightgaussian}
Z.~Fan, K.~Wang, K.~Wen, Z.~Zhu, D.~Xu, Z.~Wang, et~al.
\newblock Lightgaussian: Unbounded 3d gaussian compression with 15x reduction and 200+ fps.
\newblock {\em Advances in neural information processing systems}, 37:140138--140158, 2024.

\bibitem{kplanes}
S.~Fridovich-Keil, G.~Meanti, F.~R. Warburg, B.~Recht, and A.~Kanazawa.
\newblock K-planes: Explicit radiance fields in space, time, and appearance.
\newblock In {\em Proceedings of the IEEE/CVF Conference on Computer Vision and Pattern Recognition}, pages 12479--12488, 2023.

\bibitem{fu2022octattention}
C.~Fu, G.~Li, R.~Song, W.~Gao, and S.~Liu.
\newblock Octattention: Octree-based large-scale contexts model for point cloud compression.
\newblock In {\em Proceedings of the AAAI conference on artificial intelligence}, volume~36, pages 625--633, 2022.

\bibitem{kitti}
A.~Geiger, P.~Lenz, C.~Stiller, and R.~Urtasun.
\newblock Vision meets robotics: The kitti dataset.
\newblock {\em The international journal of robotics research}, 32(11):1231--1237, 2013.

\bibitem{gray1984vector}
R.~Gray.
\newblock Vector quantization.
\newblock {\em IEEE Assp Magazine}, 1(2):4--29, 1984.

\bibitem{hedman2018deep}
P.~Hedman, J.~Philip, T.~Price, J.-M. Frahm, G.~Drettakis, and G.~Brostow.
\newblock Deep blending for free-viewpoint image-based rendering.
\newblock {\em ACM Transactions on Graphics (ToG)}, 37(6):1--15, 2018.

\bibitem{3dgs}
B.~Kerbl, G.~Kopanas, T.~Leimk{\"u}hler, and G.~Drettakis.
\newblock 3d gaussian splatting for real-time radiance field rendering.
\newblock {\em ACM Trans. Graph.}, 42(4):139--1, 2023.

\bibitem{knapitsch2017tanks}
A.~Knapitsch, J.~Park, Q.-Y. Zhou, and V.~Koltun.
\newblock Tanks and temples: Benchmarking large-scale scene reconstruction.
\newblock {\em ACM Transactions on Graphics (ToG)}, 36(4):1--13, 2017.

\bibitem{lee2024compact}
J.~C. Lee, D.~Rho, X.~Sun, J.~H. Ko, and E.~Park.
\newblock Compact 3d gaussian representation for radiance field.
\newblock In {\em Proceedings of the IEEE/CVF Conference on Computer Vision and Pattern Recognition}, pages 21719--21728, 2024.

\bibitem{ling2024dl3dv}
L.~Ling, Y.~Sheng, Z.~Tu, W.~Zhao, C.~Xin, K.~Wan, L.~Yu, Q.~Guo, Z.~Yu, Y.~Lu, et~al.
\newblock Dl3dv-10k: A large-scale scene dataset for deep learning-based 3d vision.
\newblock In {\em Proceedings of the IEEE/CVF Conference on Computer Vision and Pattern Recognition}, pages 22160--22169, 2024.

\bibitem{liu2024hemgs}
L.~Liu, Z.~Chen, and D.~Xu.
\newblock Hemgs: A hybrid entropy model for 3d gaussian splatting data compression.
\newblock {\em arXiv preprint arXiv:2411.18473}, 2024.

\bibitem{liu2025compgs++}
X.~Liu, X.~Wu, S.~Wang, Z.~Li, and S.~Kwong.
\newblock Compgs++: Compressed gaussian splatting for static and dynamic scene representation.
\newblock {\em arXiv preprint arXiv:2504.13022}, 2025.

\bibitem{liu2024compgs}
X.~Liu, X.~Wu, P.~Zhang, S.~Wang, Z.~Li, and S.~Kwong.
\newblock Compgs: Efficient 3d scene representation via compressed gaussian splatting.
\newblock In {\em Proceedings of the 32nd ACM International Conference on Multimedia}, pages 2936--2944, 2024.

\bibitem{lu2024scaffold}
T.~Lu, M.~Yu, L.~Xu, Y.~Xiangli, L.~Wang, D.~Lin, and B.~Dai.
\newblock Scaffold-gs: Structured 3d gaussians for view-adaptive rendering.
\newblock In {\em Proceedings of the IEEE/CVF Conference on Computer Vision and Pattern Recognition}, pages 20654--20664, 2024.

\bibitem{luo2024scp}
A.~Luo, L.~Song, K.~Nonaka, K.~Unno, H.~Sun, M.~Goto, and J.~Katto.
\newblock Scp: spherical-coordinate-based learned point cloud compression.
\newblock In {\em Proceedings of the AAAI Conference on Artificial Intelligence}, volume~38, pages 3954--3962, 2024.

\bibitem{ma2025enhancing}
J.~Ma, Y.~Hu, L.~Tang, J.~Yang, Y.~Zhai, and R.~Wang.
\newblock Enhancing 3d gaussian splatting compression via spatial condition-based prediction.
\newblock {\em arXiv preprint arXiv:2503.23337}, 2025.

\bibitem{mildenhall2021nerf}
B.~Mildenhall, P.~P. Srinivasan, M.~Tancik, J.~T. Barron, R.~Ramamoorthi, and R.~Ng.
\newblock Nerf: Representing scenes as neural radiance fields for view synthesis.
\newblock {\em Communications of the ACM}, 65(1):99--106, 2021.

\bibitem{morgenstern2024compact}
W.~Morgenstern, F.~Barthel, A.~Hilsmann, and P.~Eisert.
\newblock Compact 3d scene representation via self-organizing gaussian grids.
\newblock In {\em European Conference on Computer Vision}, pages 18--34. Springer, 2024.

\bibitem{navaneet2024compgs}
K.~Navaneet, K.~Pourahmadi~Meibodi, S.~Abbasi~Koohpayegani, and H.~Pirsiavash.
\newblock Compgs: Smaller and faster gaussian splatting with vector quantization.
\newblock In {\em European Conference on Computer Vision}, pages 330--349. Springer, 2024.

\bibitem{niedermayr2024compressed}
S.~Niedermayr, J.~Stumpfegger, and R.~Westermann.
\newblock Compressed 3d gaussian splatting for accelerated novel view synthesis.
\newblock In {\em Proceedings of the IEEE/CVF Conference on Computer Vision and Pattern Recognition}, pages 10349--10358, 2024.

\bibitem{papantonakis2024reducing}
P.~Papantonakis, G.~Kopanas, B.~Kerbl, A.~Lanvin, and G.~Drettakis.
\newblock Reducing the memory footprint of 3d gaussian splatting.
\newblock {\em Proceedings of the ACM on Computer Graphics and Interactive Techniques}, 7(1):1--17, 2024.

\bibitem{song2023efficient}
R.~Song, C.~Fu, S.~Liu, and G.~Li.
\newblock Efficient hierarchical entropy model for learned point cloud compression.
\newblock In {\em Proceedings of the IEEE/CVF Conference on Computer Vision and Pattern Recognition}, pages 14368--14377, 2023.

\bibitem{tangandyang2023torchsparse}
H.~Tang, S.~Yang, Z.~Liu, K.~Hong, Z.~Yu, X.~Li, G.~Dai, Y.~Wang, and S.~Han.
\newblock Torchsparse++: Efficient training and inference framework for sparse convolution on gpus.
\newblock In {\em IEEE/ACM International Symposium on Microarchitecture (MICRO)}, 2023.

\bibitem{tang2025neuralgs}
Z.~Tang, C.~Feng, X.~Cheng, W.~Yu, J.~Zhang, Y.~Liu, X.~Long, W.~Wang, and L.~Yuan.
\newblock Neuralgs: Bridging neural fields and 3d gaussian splatting for compact 3d representations.
\newblock {\em arXiv preprint arXiv:2503.23162}, 2025.

\bibitem{wang2024end}
H.~Wang, H.~Zhu, T.~He, R.~Feng, J.~Deng, J.~Bian, and Z.~Chen.
\newblock End-to-end rate-distortion optimized 3d gaussian representation.
\newblock In {\em European Conference on Computer Vision}, pages 76--92. Springer, 2024.

\bibitem{sparsepcgc}
J.~Wang, D.~Ding, Z.~Li, X.~Feng, C.~Cao, and Z.~Ma.
\newblock Sparse tensor-based multiscale representation for point cloud geometry compression.
\newblock {\em IEEE Transactions on Pattern Analysis and Machine Intelligence}, 2022.

\bibitem{unipcgc}
K.~Wang and W.~Gao.
\newblock Unipcgc: Towards practical point cloud geometry compression via an efficient unified approach.
\newblock {\em Proceedings of the AAAI Conference on Artificial Intelligence}, 39(12):12721--12729, Apr. 2025.

\bibitem{wang2025tc}
T.~Wang, Z.~Yu, and Y.~Xu.
\newblock Tc-gs: Tri-plane based compression for 3d gaussian splatting.
\newblock {\em arXiv preprint arXiv:2503.20221}, 2025.

\bibitem{wang2024contextgs}
Y.~Wang, Z.~Li, L.~Guo, W.~Yang, A.~Kot, and B.~Wen.
\newblock Contextgs: Compact 3d gaussian splatting with anchor level context model.
\newblock {\em Advances in neural information processing systems}, 37:51532--51551, 2024.

\bibitem{xie2024sizegs}
S.~Xie, J.~Liu, W.~Zhang, S.~Ge, S.~Pan, C.~Tang, Y.~Bai, and Z.~Wang.
\newblock Sizegs: Size-aware compression of 3d gaussians with hierarchical mixed precision quantization.
\newblock {\em arXiv preprint arXiv:2412.05808}, 2024.

\bibitem{xie2024mesongs}
S.~Xie, W.~Zhang, C.~Tang, Y.~Bai, R.~Lu, S.~Ge, and Z.~Wang.
\newblock Mesongs: Post-training compression of 3d gaussians via efficient attribute transformation.
\newblock In {\em European Conference on Computer Vision}, pages 434--452. Springer, 2024.

\bibitem{xu2024fakeshield}
Z.~Xu, X.~Zhang, R.~Li, Z.~Tang, Q.~Huang, and J.~Zhang.
\newblock Fakeshield: Explainable image forgery detection and localization via multi-modal large language models.
\newblock {\em arXiv preprint arXiv:2410.02761}, 2024.

\bibitem{you2025reno}
K.~You, T.~Chen, D.~Ding, M.~S. Asif, and Z.~Ma.
\newblock Reno: Real-time neural compression for 3d lidar point clouds.
\newblock {\em arXiv preprint arXiv:2503.12382}, 2025.

\bibitem{zhancat}
Y.-T. Zhan, C.-Y. Ho, H.~Yang, Y.-H. Chen, J.~C. Chiang, Y.-L. Liu, and W.-H. Peng.
\newblock Cat-3dgs: A context-adaptive triplane approach to rate-distortion-optimized 3dgs compression.
\newblock In {\em The Thirteenth International Conference on Learning Representations}.

\bibitem{zhan2025cat}
Y.-T. Zhan, H.-b. Yang, C.-Y. Ho, J.-C. Chiang, and W.-H. Peng.
\newblock Cat-3dgs pro: A new benchmark for efficient 3dgs compression.
\newblock {\em arXiv preprint arXiv:2503.12862}, 2025.

\bibitem{zhang2025adadpcc}
C.~Zhang and W.~Gao.
\newblock Adadpcc: Adaptive rate control and rate-distortion-complexity optimization for dynamic point cloud compression.
\newblock In {\em Proceedings of the AAAI Conference on Artificial Intelligence}, volume~39, pages 13188--13196, 2025.

\bibitem{zhang2024gs}
X.~Zhang, J.~Meng, R.~Li, Z.~Xu, Y.~Zhang, and J.~Zhang.
\newblock Gs-hider: Hiding messages into 3d gaussian splatting.
\newblock {\em arXiv preprint arXiv:2405.15118}, 2024.

\end{thebibliography}
}

\newpage
\appendix
\section*{Technical Appendices}


\section{GausPcgc Framework}
\label{app:gauspcc_method}
\subsection{Detailed Description}
\begin{figure}
    \centering
    \includegraphics[width=1.0\linewidth]{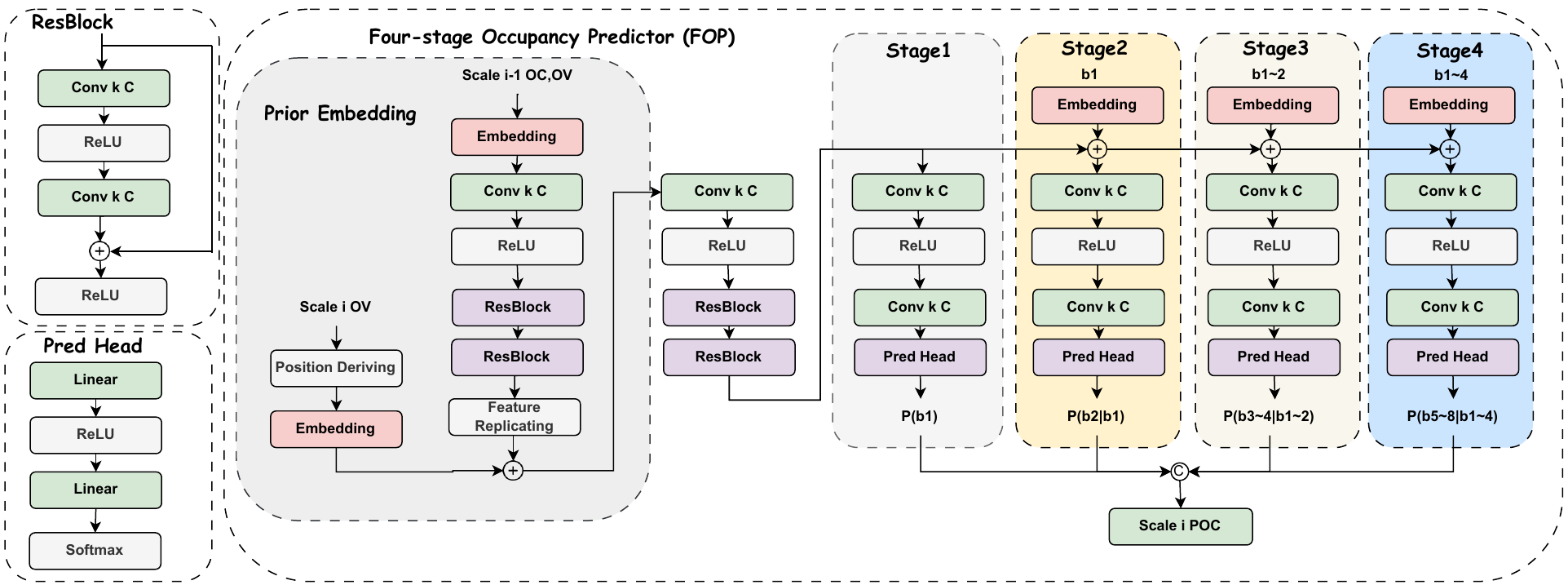}
     \caption{
     Detailed framework diagram of Four-stage Occupancy Predictor.
     }
    \label{fig:fop}
\end{figure}

We present the detailed architecture of our Four-stage Occupancy Predictor (FOP). All convolutions described in this section refer to sparse convolutions implemented with Torchsparse \cite{tangandyang2023torchsparse}. As illustrated in Figure \ref{fig:fop}, the Prior Embedding module extracts features from available prior information, similar to RENO \cite{you2025reno}. Following feature extraction, we employ a serial four-stage occupancy code prediction scheme. We adopt a non-uniform grouping strategy, dividing the 8-bit occupancy code into 1-1-2-4 segments for prediction. This approach ensures that each encoding step can reference prior information with matching bit precision. Additionally, before each occupancy probability prediction, we utilize stacked Conv-ReLU-Conv blocks (Spatial convolution) to aggregate neighboring prior features, enabling more accurate occupancy probability estimation. Throughout the FOP architecture, all convolutions use kernel size $k=5$ with channel dimension $C=32$.

\subsection{Pseudocode}
\begin{algorithm}
\caption{GausPcgc for Point Cloud Compression}
\label{algor:gauspcc}
\begin{algorithmic}[1]
\Require Point cloud $\mathcal{P}$, voxel size $\tau$
\Ensure Compressed bitstream

\State $\{\mathcal{X}^{(l)}\}_{l=0}^L \gets \text{VoxelizeMultiScale}(\mathcal{P}, \tau)$ \Comment{Multi-scale voxelization}
\State $\text{total\_bits} \gets 0$
\State $N \gets |\mathcal{P}|$ \Comment{Number of points in the point cloud}

\For{$i = L-1$ \textbf{downto} $0$} \Comment{Process from coarse to fine scales}
    \State $\text{OC}_{i-1} \gets$ Occupancy codes at scale $i-1$
    \State $\mathcal{C}_{i} \gets$ Coordinates at scale $i$
    \State $\mathbf{F}_{i} \gets \text{FeatureExtractor}(\text{OC}_{i-1}, \mathcal{C}_{i})$ \Comment{Extract initial features}
    
    \For{each voxel $v$ at scale $i$}
        \State $\text{oc} \gets \text{OC}_{i}[v]$ \Comment{Ground truth occupancy code}
        
        \State \textcolor{blue}{// Stage 1: Predict first bit}
        \State $b_1 \gets \lfloor \text{oc} / 128 \rfloor \bmod 2$
        \State $\mathbf{F}_{v,1} \gets \mathcal{S}_1(\mathbf{F}_{i}[v])$ \Comment{Apply spatial convolution}
        \State $P(b_1|\mathcal{C}_i) \gets \sigma(\mathcal{H}_1(\mathbf{F}_{v,1}))$ \Comment{Predict probability}
        \State $\text{bits}_1 \gets -\log_2(P(b_1|\mathcal{C}_i) + \epsilon)$ \Comment{Calculate entropy}
        
        \State \textcolor{blue}{// Stage 2: Predict second bit}
        \State $b_2 \gets \lfloor \text{oc} / 64 \rfloor \bmod 2$
        \State $\mathbf{F}_{v,2} \gets \mathcal{S}_2(\mathbf{F}_{i}[v] + \mathcal{E}_1(b_1))$ \Comment{Embed previous bit}
        \State $P(b_2|b_1,\mathcal{C}_i) \gets \sigma(\mathcal{H}_2(\mathbf{F}_{v,2}))$
        \State $\text{bits}_2 \gets -\log_2(P(b_2|b_1,\mathcal{C}_i) + \epsilon)$
        
        \State \textcolor{blue}{// Stage 3: Predict third and fourth bits}
        \State $b_{3:4} \gets \lfloor \text{oc} / 16 \rfloor \bmod 4$
        \State $\mathbf{F}_{v,3} \gets \mathcal{S}_3(\mathbf{F}_{i}[v] + \mathcal{E}_2(2b_1 + b_2))$
        \State $P(b_{3:4}|b_1,b_2,\mathcal{C}_i) \gets \sigma(\mathcal{H}_3(\mathbf{F}_{v,3}))$
        \State $\text{bits}_3 \gets -\log_2(P(b_{3:4}|b_1,b_2,\mathcal{C}_i) + \epsilon)$
        
        \State \textcolor{blue}{// Stage 4: Predict remaining four bits}
        \State $b_{5:8} \gets \text{oc} \bmod 16$
        \State $\mathbf{F}_{v,4} \gets \mathcal{S}_4(\mathbf{F}_{i}[v] + \mathcal{E}_3(4(2b_1 + b_2) + b_{3:4}))$
        \State $P(b_{5:8}|b_1,b_2,b_{3:4},\mathcal{C}_i) \gets \sigma(\mathcal{H}_4(\mathbf{F}_{v,4}))$
        \State $\text{bits}_4 \gets -\log_2(P(b_{5:8}|b_1,b_2,b_{3:4},\mathcal{C}_i) + \epsilon)$
        
        \State $\text{total\_bits} \gets \text{total\_bits} + \text{bits}_1 + \text{bits}_2 + \text{bits}_3 + \text{bits}_4$
    \EndFor
\EndFor

\State $\text{BPP} \gets \text{total\_bits} / N$ \Comment{Bits per point}
\State \Return $\text{ArithmeticEncode}(\{\mathcal{X}^{(l)}\}_{l=0}^L, \text{predicted probabilities})$
\end{algorithmic}
\end{algorithm}
We describe the overall point cloud compression pipeline of GausPcgc using pseudocode, as shown in Algorithm \ref{algor:gauspcc}.

\section{More Results}
\subsection{Current Point Cloud Compression Result Analysis}
\label{app:sub:current_ana}
On the Kitti\cite{kitti} dataset, the compression performance of G-PCC\cite{gpcc} is obtained from our own reproduction based on the official implementation, while the results of SparsePCGC\cite{sparsepcgc}, Octattention\cite{fu2022octattention}, EHEM\cite{song2023efficient}, and RENO\cite{you2025reno} are taken from the RD curves published in their respective original papers. For a fair comparison, we select the bitrate points corresponding to the same PSNR level under 14-bit quantization precision on the Kitti dataset from the RD curves of each method. On the 8iVFB dataset, the compression performance of G-PCC is similarly obtained from our own reproduction. The results of SparsePCGC, Octattention, and EHEM are referenced from the RD curves reported in their original papers. Since RENO does not provide results on the 8iVFB dataset, we reproduce its performance following the same training and evaluation protocols used for the other methods. The results are summarized in the table below.

\begin{table}[t]
\centering
\caption{Compression performance comparison on Kitti and 8iVFB datasets}
\begin{tabular}{l|c|c|c|c|c}
\toprule
        & G-PCC & SparsePCGC & Octattention & EHEM & RENO \\
\midrule
Kitti (bpp)   & 8.83  & 6.59       & 7.11         & 5.31    & 7.51    \\
CR-Gain   & 0.00\%   & -25.4\%      & -19.5\%        & -39.9\%   & -14.9\%   \\ \toprule
8iVFB (bpp)   & 0.76  & 0.59       & 0.68         & 0.64    & 0.70    \\
CR-Gain   & 0.00\%   & -22.4\%      & -10.5\%        & -15.8\%   & -7.9\%    \\
\bottomrule
\end{tabular}
\label{tab:compression_comparison}
\end{table}

As shown in Table \ref{tab:compression_comparison}, EHEM achieves the best compression performance on the Kitti dataset, yielding a 39.9\% bitrate reduction compared to G-PCC, demonstrating strong adaptability to conventional LiDAR point clouds. On the 8iVFB dataset, SparsePCGC performs the best, achieving a 22.4\% bitrate reduction relative to G-PCC. These results indicate that different methods exhibit varying strengths depending on the type of point cloud data, and that compression performance is highly influenced by the underlying data distribution. Therefore, when designing compression models, it is important to consider the characteristics and structural properties of the target dataset to achieve more effective and tailored optimization.

\subsection{Gaussian Point Cloud Compression Result Analysis}
\label{app:sub: gaussian_ana}

Table \ref{table:pcc_res} presents the results of our established Gaussian point cloud compression benchmark. Among the evaluated methods, RENO \cite{you2025reno} and SparsePCGC \cite{sparsepcgc} employ voxel-based hierarchical prediction, while EHEM \cite{song2023efficient} and Octattention \cite{fu2022octattention} utilize octree-structured hierarchical prediction. A joint analysis of Table \ref{table:pcc_res} and Table \ref{tab:compression_comparison} reveals an interesting phenomenon: on Kitti, EHEM significantly outperforms Octattention, while on 8iVFB, SparsePCGC surpasses RENO. However, the results on Gaussian data are completely opposite. This further proves that optimizations designed for traditional point cloud datasets are not suitable for Gaussian point clouds. This aligns with the local density analysis and fractal dimension analysis presented earlier, indicating that the optimization methods proposed in previous works lose their performance gains due to changes in point cloud distribution. Next will provide a detailed explanation from the perspective of network design.

\textbf{RENO and SparsePCGC.} RENO replaces SparsePCGC's upsample-then-prune encoding approach with octree occupancy codes. RENO benefits from performing convolution operations at lower resolution layers, thereby obtaining larger receptive fields. In contrast, SparsePCGC doubles the resolution after upsampling, expanding coordinate intervals, which disadvantages sparse convolution operations with limited receptive fields. Furthermore, SparsePCGC's 8-stage grouping essentially leverages spatial-level prior information, whereas RENO utilizes channel-level priors for occupancy codes. In the Gaussian point cloud distribution characterized by local density and global sparsity, spatial-level priors provide less prior information in sparse regions because the points are farther apart with lower correlation. In contrast, channel-level priors can offer more stable prior information across both sparse and dense areas, enabling RENO to achieve better performance.


\textbf{EHEM and Octattention.} EHEM and Octattention both employ octree-based approaches, requiring octree construction before processing, which introduces additional time and space complexity compared to RENO. During decoding, real-time construction of decoding octrees imposes significant CPU overhead. Octattention achieves optimal rate performance in our experiments through autoregressive methods, but its excessive decoding time due to autoregression is impractical. EHEM, an improved version of Octattention, addresses the prohibitive decoding time by implementing two-stage checkerboard spatial-level prior information for encoding and decoding. However, as previously analyzed, spatial-level priors struggle to provide additional information in highly sparse Gaussian point clouds. This explains why the grid-based strategies, effective on LiDAR point clouds, do not perform well on Gaussian point clouds. Additionally, the DGCNN neighborhood feature extraction in EHEM fails to deliver the same performance improvements on irregularly distributed Gaussian point clouds as it does with LiDAR data. These phenomena indicate that the existing enhancement strategies, developed and tested on previous point cloud datasets such as Kitti, 8iVFB, etc., are not well-suited for Gaussian point clouds with their unique distribution characteristics.


\subsection{Rendering Visualization}
\begin{figure}[t]
    \centering
    \begin{subfigure}[b]{0.48\textwidth}
        \centering
        \includegraphics[width=\textwidth]{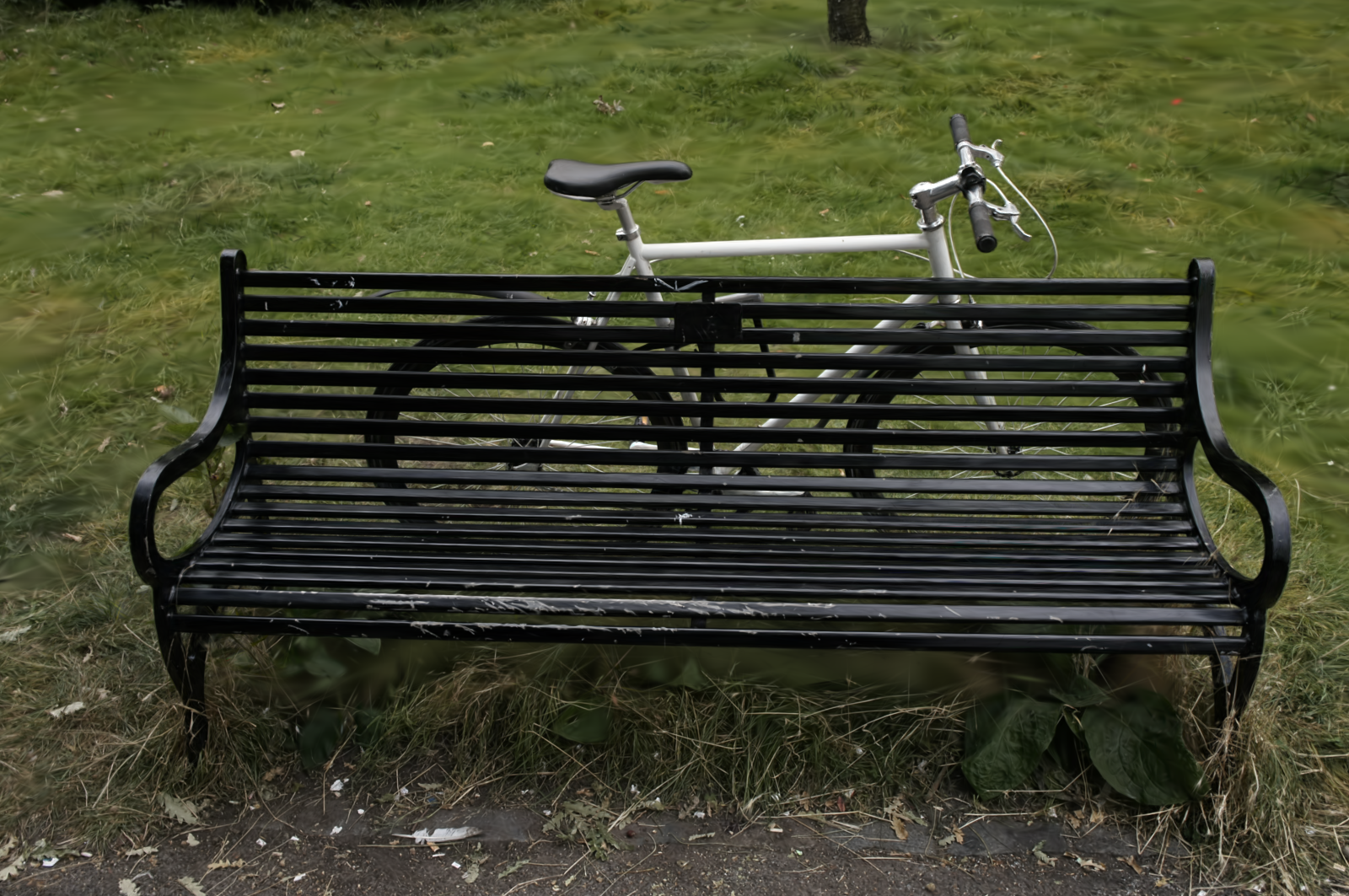}
        \caption{3DGS PSNR: 26.99dB Size: 1126.2MB}
        \label{fig:3dgs-00015}
    \end{subfigure}
    \hfill 
    \begin{subfigure}[b]{0.48\textwidth}
        \centering
        \includegraphics[width=\textwidth]{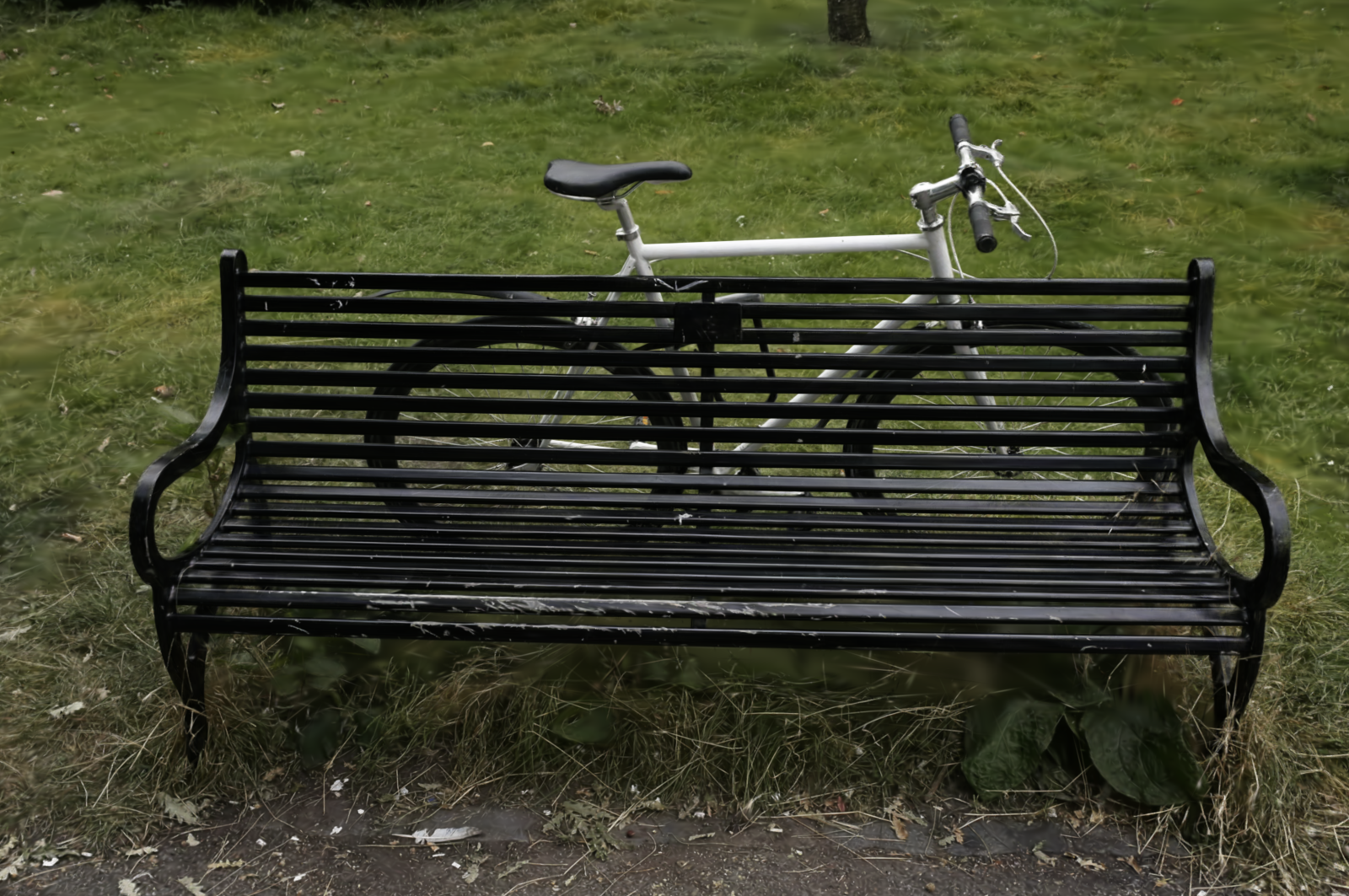}
        \caption{Scaffold-GS PSNR: 27.31dB Size: 271.16MB}
        \label{fig:scaf-00015}
    \end{subfigure}
    
    \vspace{0.5cm} 
    
    \begin{subfigure}[b]{0.48\textwidth}
        \centering
        \includegraphics[width=\textwidth]{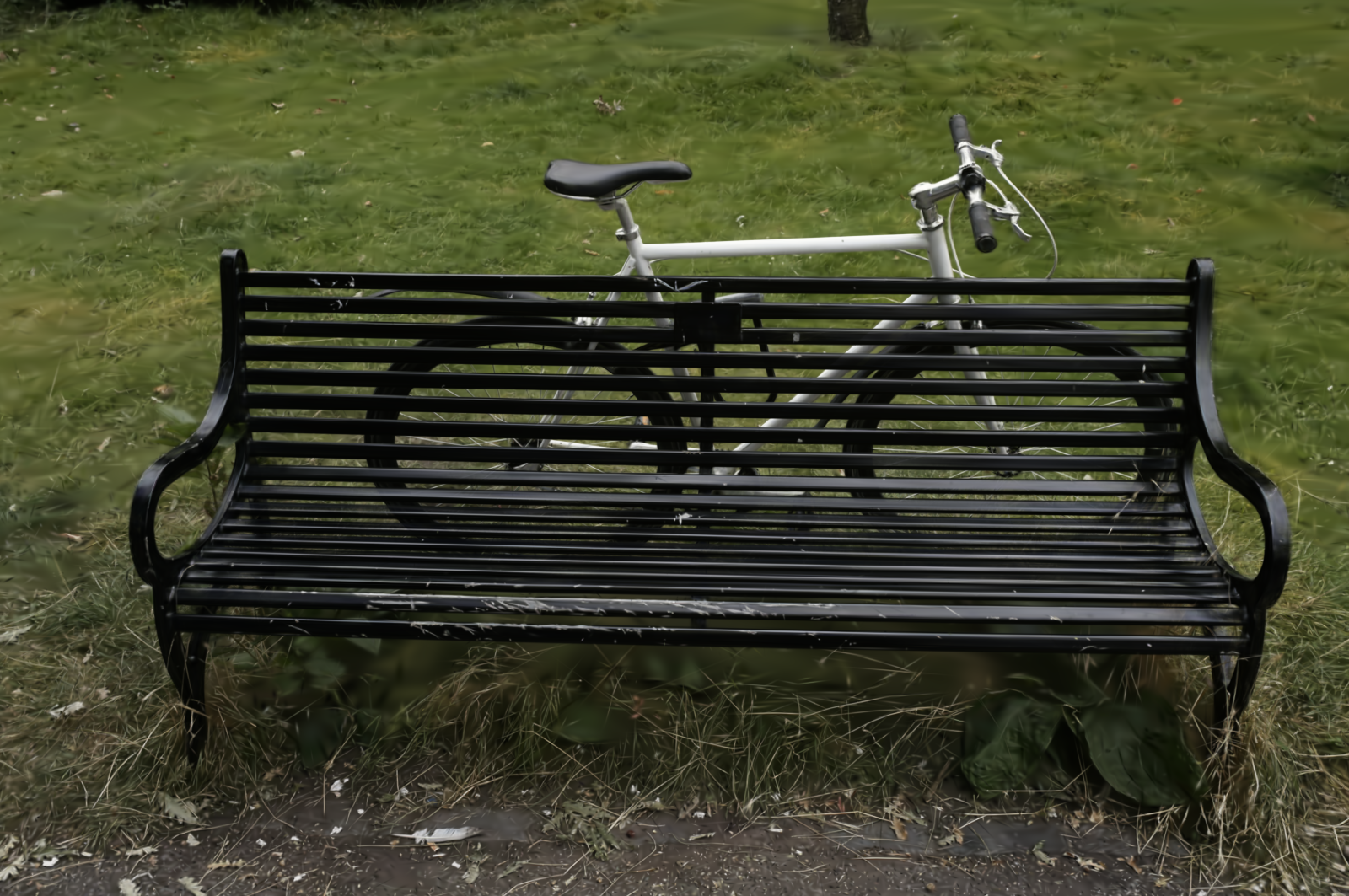}
        \caption{Ours-TC-GS PSNR: 27.04dB Size: 16.32MB}
        \label{fig: our-tcgs-00015}
    \end{subfigure}
    \hfill  
    \begin{subfigure}[b]{0.48\textwidth}
        \centering
        \includegraphics[width=\textwidth]{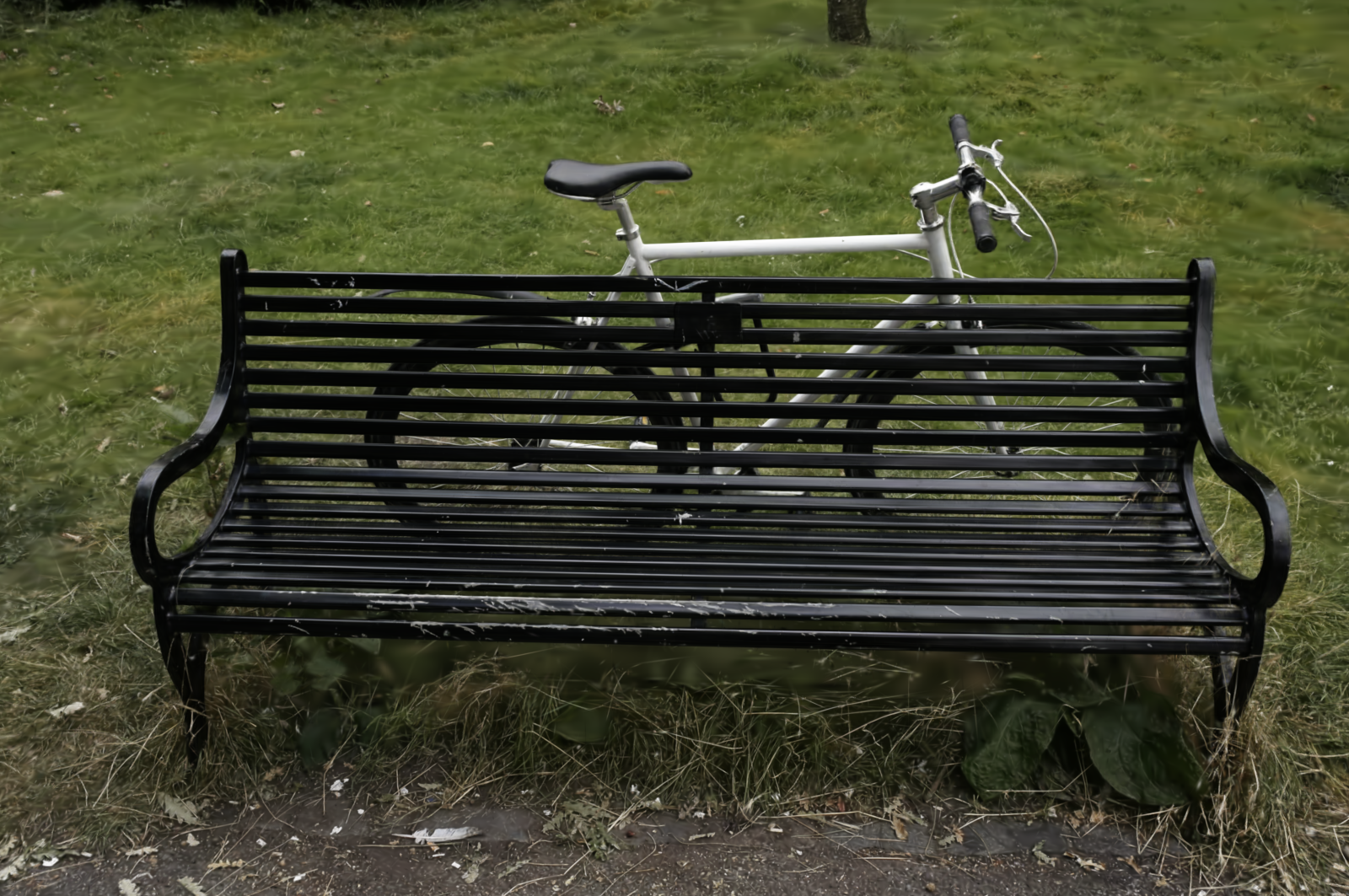}
        \caption{Ours-HAC PSNR: 27.44dB Size: 36.39MB }
        \label{fig:our-hac-00015}
    \end{subfigure}
    \caption{Visual comparison of rendering results on the Mip-NeRF 360 \cite{barron2022mipnerf360} bicycle scene. We showcase two widely-adopted baseline methods alongside two variants of our proposed approach.}
    \label{fig:bicycle_vis}
     \vspace{-8pt}
\end{figure}

\begin{figure}[htbp]
    \centering
    \begin{subfigure}[b]{0.48\textwidth}
        \centering
        \includegraphics[width=\textwidth]{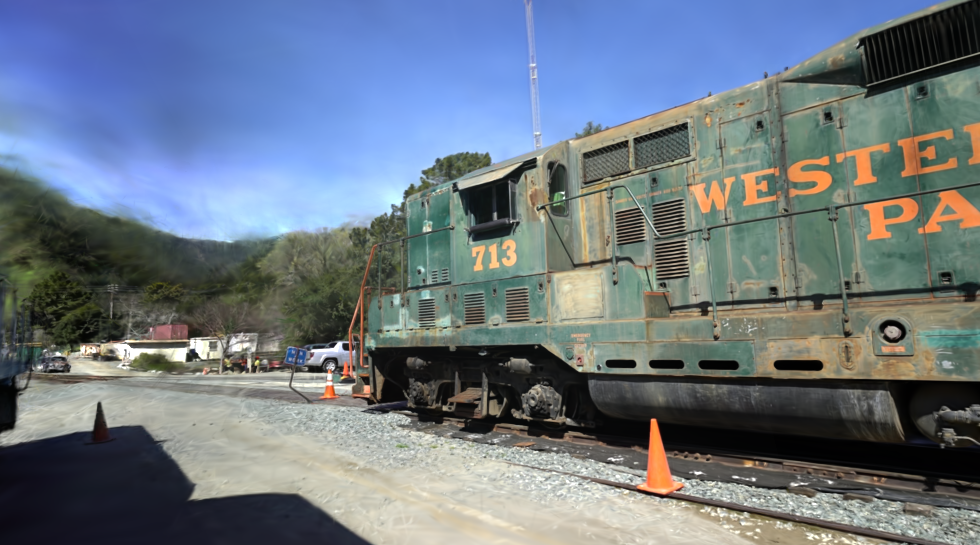}
        \caption{3DGS PSNR: 22.03dB Size: 257.3MB}
        \label{fig:3dgs-00004}
    \end{subfigure}
    \hfill 
    \begin{subfigure}[b]{0.48\textwidth}
        \centering
        \includegraphics[width=\textwidth]{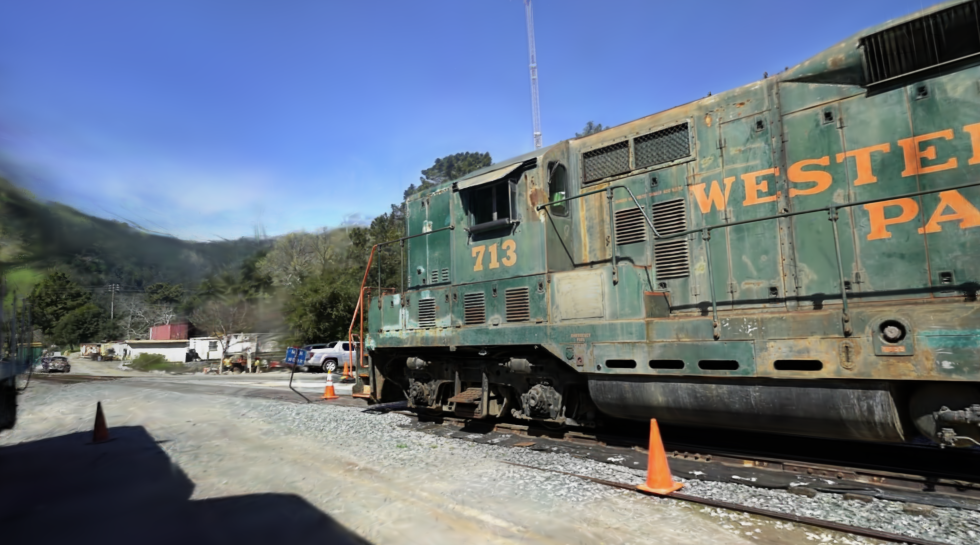}
        \caption{Scaffold-GS PSNR: 22.37dB Size: 92.24MB}
        \label{fig:scaf-00004}
    \end{subfigure}
    
    \vspace{0.5cm}  
    
    \begin{subfigure}[b]{0.48\textwidth}
        \centering
        \includegraphics[width=\textwidth]{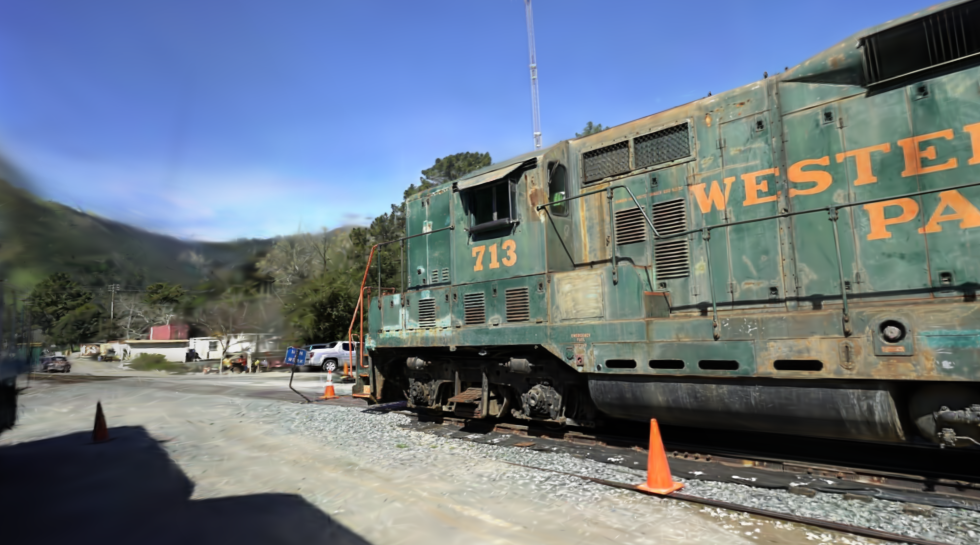}
        \caption{Ours-TC-GS PSNR: 22.15dB Size: 6.22MB}
        \label{fig: our-tcgs-00004}
    \end{subfigure}
    \hfill 
    \begin{subfigure}[b]{0.48\textwidth}
        \centering
        \includegraphics[width=\textwidth]{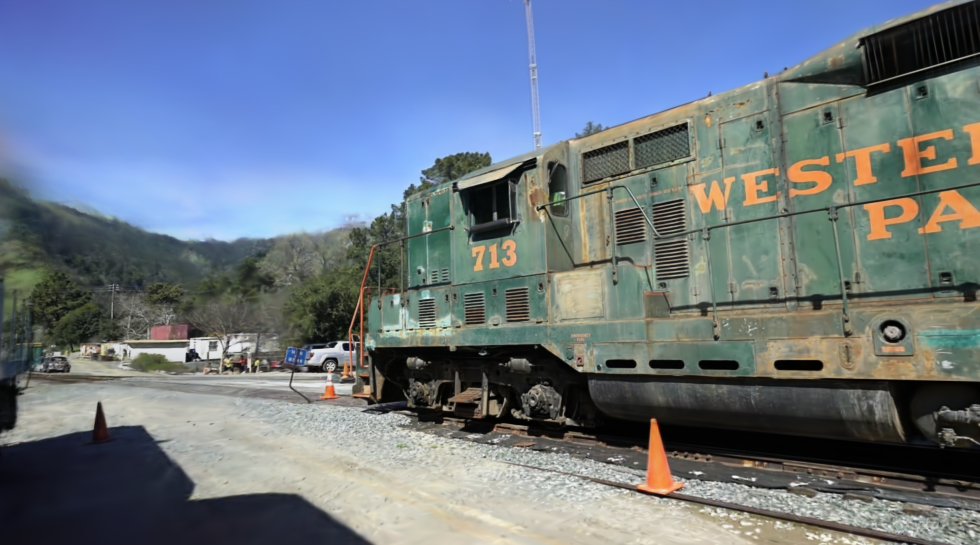}
        \caption{Ours-HAC PSNR: 22.94dB Size: 9.17MB }
        \label{fig:our-hac-00004}
    \end{subfigure}
    \caption{Visual comparison of rendering results on the Tanks and Temples \cite{knapitsch2017tanks} train scene. We showcase two widely-adopted baseline methods alongside two variants of our proposed approach.}
    \label{fig:train_vis}
\end{figure}

We present visualization of rendering results from our methods compared with baseline approaches, as shown in Figure \ref{fig:bicycle_vis} and Figure \ref{fig:train_vis}. As clearly demonstrated in these visualizations, our proposed methods achieve remarkable compression rates while preserving or even enhancing visual quality. Specifically, in the challenging bicycle scene from the Mip-NeRF 360 \cite{barron2022mipnerf360} dataset and the complex train scene from Tanks and Temples \cite{knapitsch2017tanks} collection, our GausPcgc-TC-GS method accomplishes substantial compression rates of 69× and 41× respectively, all while maintaining visual fidelity comparable to the original 3DGS renderings. Furthermore, our advanced GausPcgc-HAC approach delivers even more impressive results, achieving compression rates of 31× and 28× for these scenes while simultaneously providing enhanced visual quality with better preservation of fine details and improved overall appearance. These results convincingly demonstrate the effectiveness of our compression framework in balancing high compression efficiency with exceptional visual performance.

\subsection{RD curve}
\begin{figure}
    \centering
    \includegraphics[width=1.0\linewidth]{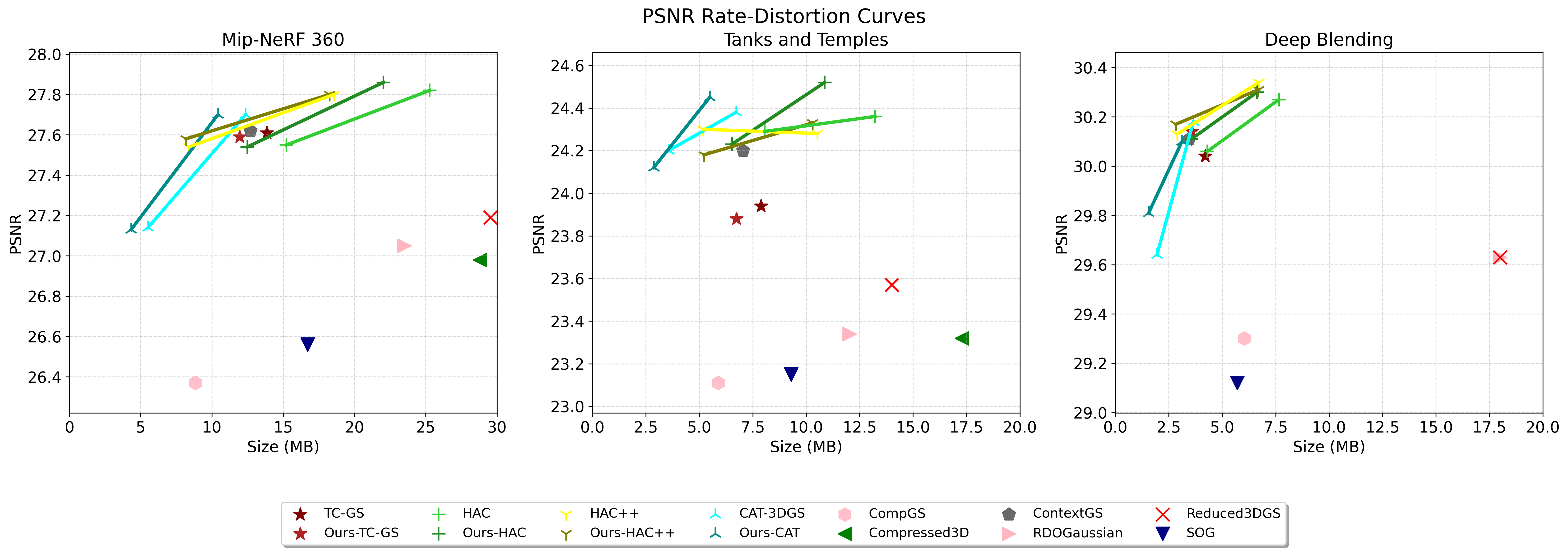}
    \includegraphics[width=1.0\linewidth]{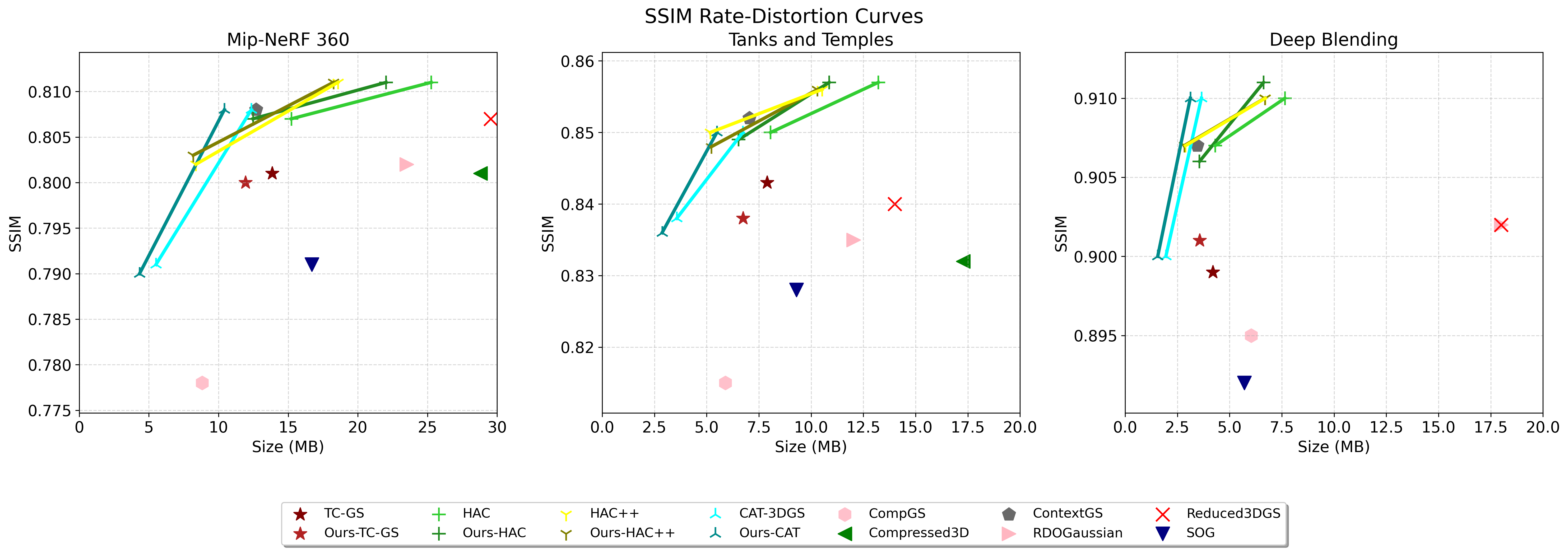}
    \includegraphics[width=1.0\linewidth]{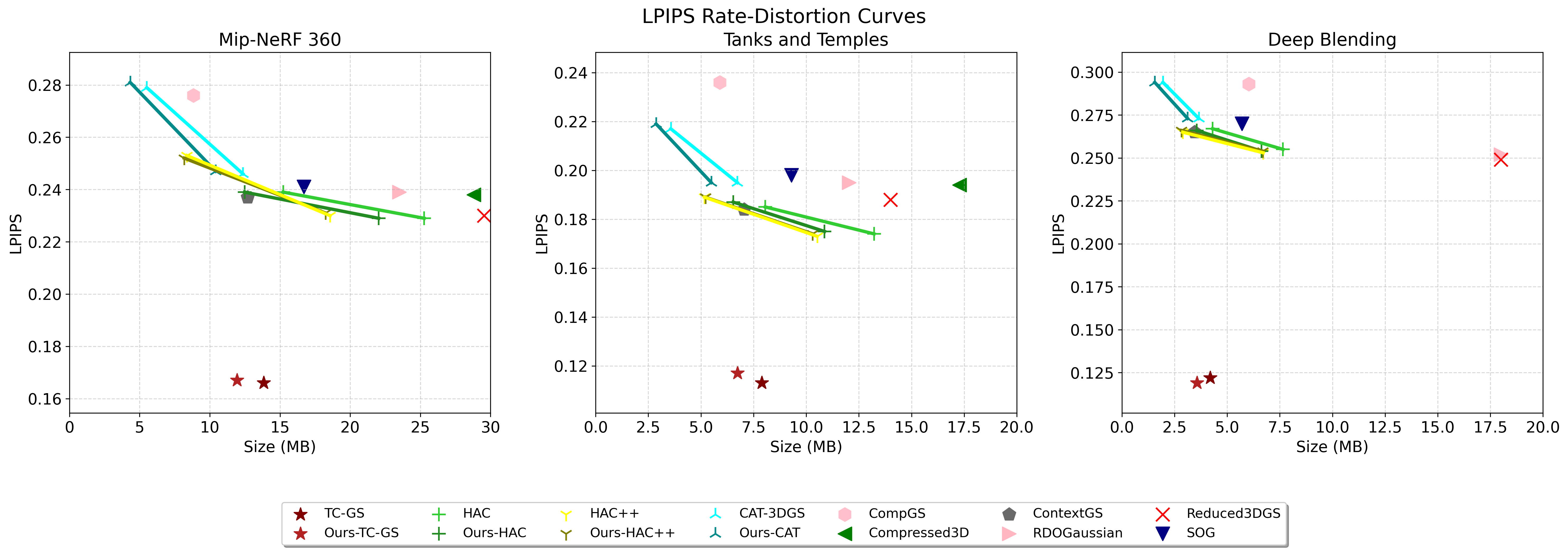}
     \caption{
     RD curves of various methods. For better visualization, we select only the top 14 methods for plotting. The metrics used include PSNR, SSIM, and LPIPS.
     }
    \label{fig:rd_res}
\end{figure}

As shown in Figure \ref{fig:rd_res}, we present the RD curves of the top 14 methods. The results clearly demonstrate that methods utilizing GausPcgc as a Gaussian geometry compressor achieve consistent performance improvements. As an efficient geometry compressor, GausPcgc-enhanced methods evidently offer greater advantages in terms of model size compared to previous approaches. Moreover, our method exhibits superior PSNR performance on certain datasets. This improvement largely stems from the fact that methods like HAC \cite{chen2024hac} and Cat-3DGS \cite{zhancat} employ 16-bit quantization when processing geometric positions, which can potentially cause point duplication, resulting in lossy geometry representation. In Scaffold-GS \cite{lu2024scaffold} based methods, the geometric information is inherently structured, allowing quantization based solely on voxel size to ensure strictly lossless quantization. GausPcgc adopts voxel size-based quantization, thereby guaranteeing strictly lossless geometric representation.

\subsection{Bitstream Size Analysis}
\label{app:sub:bits ana}
We analyze the bitstream size of each component in our method and its enhanced variants, as shown in Table \ref{table:size_hac}, \ref{table:size_cat}, and \ref{table:size_tcgs}. The results demonstrate that our approach not only significantly compresses the Gaussian positions, but also achieves lossless compression by adopting a quantization strategy that maintains the same granularity as the voxel size. Consequently, our method requires fewer anchors to represent the entire scene, resulting in a substantial reduction in anchor count.
\begin{table}[]
\caption{Detailed bitstream analysis for the Mip-NeRF 360 \cite{barron2022mipnerf360} bicycle scene. We present the specific bit allocation for each component, along with the number of anchors and visual fidelity of the scene. The comparison includes Cat-3DGS \cite{zhancat} and our proposed Ours-Cat-3DGS method.}
\label{table:size_cat}
\centering
\resizebox{1.0\textwidth}{!}{
\setlength{\tabcolsep}{3pt} 

\begin{tabular}{c|c|cccccccc|cc}
\toprule
\multirow{2}{*}{\textbf{}} & \multirow{2}{*}{\begin{tabular}[c]{@{}c@{}}Number of \\ Anchors\end{tabular}} & \multicolumn{8}{c|}{Storage Cost   (MB)}                                  & \multicolumn{2}{c}{Fidelity} \\ \cline{3-12} 
                           &                                                                               & Position & Feat    & Scaling & Offsets & Triplane\_f
   & Masks  & MLPs   & Total   & PSNR          & SSIM         \\ \toprule
Cat-3DGS                        & 623483                                                                        & 3.57
 & 10.32 & 3.00  & 3.19 & 0.18 & 0.55 & 0.35 & 21.15 & 25.21   & 0.740        \\
Ours-Cat-3DGS                   & 601928                                                                        & 0.82
 & 9.13
 & 2.84
  & 2.98
  & 0.15
  & 0.53
 & 0.35
 & 16.79
 & 25.17
         & 0.738
        \\ \toprule
\end{tabular}
}
\end{table}
\begin{table}[]
\caption{Detailed bitstream analysis for the Mip-NeRF 360 \cite{barron2022mipnerf360} bicycle scene. We present the specific bit allocation for each component, along with the number of anchors and visual fidelity of the scene. The comparison includes TC-GS \cite{wang2025tc} and our proposed Ours-TC-GS method.}
\label{table:size_tcgs}
\centering
\resizebox{1.0\textwidth}{!}{
\setlength{\tabcolsep}{3pt} 

\begin{tabular}{c|c|cccccccc|cc}
\toprule
\multirow{2}{*}{\textbf{}} & \multirow{2}{*}{\begin{tabular}[c]{@{}c@{}}Number of \\ Anchors\end{tabular}} & \multicolumn{8}{c|}{Storage Cost   (MB)}                                  & \multicolumn{2}{c}{Fidelity} \\ \cline{3-12} 
                           &                                                                               & Position & Feat    & Scaling & Offsets & Hash   & Masks  & MLPs   & Total   & PSNR          & SSIM         \\ \toprule
TC-GS                        & 511792                                                                        & 2.93 & 12.57  & 2.68  & 2.41 & 0.002 & 0.39 & 0.35 & 21.33 & 24.82         & 0.723
        \\
Ours-TC-GS                   & 450765                                                                        & 0.66
 & 10.60
 & 2.34
  & 2.04
  & 0.002  & 0.34
 & 0.35
 & 16.32
 & 24.88
         & 0.720        \\ \toprule
\end{tabular}
}
\end{table}

\section{Point Cloud Density Analysis}
In this section, we present the detailed process of point cloud local density analysis and Fractal Dimension analysis. The 8iVFB dataset is already quantized at 10-bit, so we apply no additional processing to it. For all other point clouds---including Kitti, GausPcc, and Scannet---we uniformly quantize using \texttt{voxel\_size=0.001}.
\subsection{Local Density Analysis}
\label{Local Density Analysis}

\textbf{Point cloud voxelization and discrete representation.} Given the original point cloud $\mathcal{P} = \{p_i \in \mathbb{R}^3 | i=1,2,...,N\}$, we map it to a discrete voxel space through the voxelization process $\mathcal{V}$:
\begin{equation}
\mathcal{V}: \mathbb{R}^3 \rightarrow \mathbb{Z}^3, \quad v_i = \mathcal{V}(p_i) = \lfloor p_i / \tau \rceil,
\end{equation}
where $\tau$ is the voxel size parameter, and $\lfloor \cdot \rceil$ represents the rounding operation. The resulting discrete point set after voxelization is denoted as $\mathcal{X} = \{v_i \in \mathbb{Z}^3 | i=1,2,...,M\}$, where $M \leq N$ due to the possibility of multiple points mapping to the same voxel.

\textbf{Multi-scale Representation Construction.}
A hierarchical multi-scale representation $\{\mathcal{X}^{(l)}\}_{l=0}^L$ is constructed through recursive downsampling operations, where $\mathcal{X}^{(0)} = \mathcal{X}$ represents the finest granularity voxel set:
\begin{equation}
\mathcal{X}^{(l+1)} = \mathcal{D}(\mathcal{X}^{(l)}), \quad l = 0,1,...,L-1.
\label{eq:voxelize}
\end{equation}
The downsampling operator $\mathcal{D}$ is implemented using convolution with a stride of 2:
\begin{equation}
\mathcal{D}(\mathcal{X}^{(l)}) = \{v^{(l+1)}_j \in \mathbb{Z}^3 | v^{(l+1)}_j = \lfloor v^{(l)}_i/2 \rceil, v^{(l)}_i \in \mathcal{X}^{(l)}\}.
\label{eq:downsample}
\end{equation}

\textbf{$k$-Neighborhood Voxel Count Calculation.}
For the voxel set $\mathcal{X}^{(l)}$ at each scale $l$, the $k$-neighborhood of voxel $v^{(l)}_i$ is defined as:
\begin{equation}
\mathcal{N}_k(v^{(l)}_i) = \{u \in \mathbb{Z}^3 | \|u - v^{(l)}_i\|_{\infty} \leq k/2\},
\label{eq:neighborhood}
\end{equation}
where $\|\cdot\|_{\infty}$ denotes the infinity norm, and $k$ is an odd number representing the size of the neighborhood kernel.

The neighborhood voxel count for each voxel $v^{(l)}_i$ can be computed through a convolution operation:
\begin{equation}
N(v^{(l)}_i) = (\mathcal{X}^{(l)} * \mathcal{K}_k)(v^{(l)}_i) = \sum_{u \in \mathcal{N}_k(v^{(l)}_i)} \mathbb{I}_{\mathcal{X}^{(l)}}(u),
\end{equation}
where the convolution kernel $\mathcal{K}_k \in \mathbb{R}^{k \times k \times k}$ has all elements equal to 1, and $\mathbb{I}_{\mathcal{X}^{(l)}}(\cdot)$ is an indicator function taking the value 1 when $u \in \mathcal{X}^{(l)}$ and 0 otherwise.

\textbf{Probability Density Function Estimation.}
For the set of neighborhood counts $\{N(v^{(l)}_i) | v^{(l)}_i \in \mathcal{X}^{(l)}\}$ at scale $l$, we estimate its probability density function $p^{(l)}(n)$.
First, we construct a histogram:
\begin{equation}
h^{(l)}(b_j) = \frac{|\{v^{(l)}_i | N(v^{(l)}_i) \in [b_j, b_{j+1}), v^{(l)}_i \in \mathcal{X}^{(l)}\}|}{|\mathcal{X}^{(l)}| \cdot (b_{j+1} - b_j)},
\label{eq:histogram}
\end{equation}
where $\{b_j\}_{j=0}^B$ represents the $B$ interval boundaries of the histogram, satisfying $b_0 = 0$ and $b_B = k^3$.
The histogram normalization ensures:
\begin{equation}
\int_0^{k^3} p^{(l)}(n) dn = \sum_{j=0}^{B-1} h^{(l)}(b_j) \cdot (b_{j+1} - b_j) = 1.
\end{equation}


\textbf{Multi-dataset Distribution Comparison.}
For $D$ different datasets $\{\mathcal{P}_d\}_{d=1}^D$, we can estimate the neighborhood distribution density function $\{\hat{p}_d^{(l)}(n)\}_{d=1}^D$ for each dataset at the same scale $l$, enabling quantitative comparison of distribution characteristics.

The distribution differences between datasets can be measured using statistical metrics such as Kullback-Leibler divergence:
\begin{equation}
D_{KL}(\hat{p}_i^{(l)} \| \hat{p}_j^{(l)}) = \int_0^{k^3} \hat{p}_i^{(l)}(n) \log \frac{\hat{p}_i^{(l)}(n)}{\hat{p}_j^{(l)}(n)} dn.
\label{eq:kldiv}
\end{equation}
The KL Comparison result is shown in Table \ref{table:kl}. These density functions characterize the local geometric distribution properties of different point cloud datasets at multiple hierarchical scales, providing a theoretical foundation for point cloud compression, reconstruction, and feature extraction.

\begin{figure}
    \centering
    \includegraphics[width=0.9\linewidth]{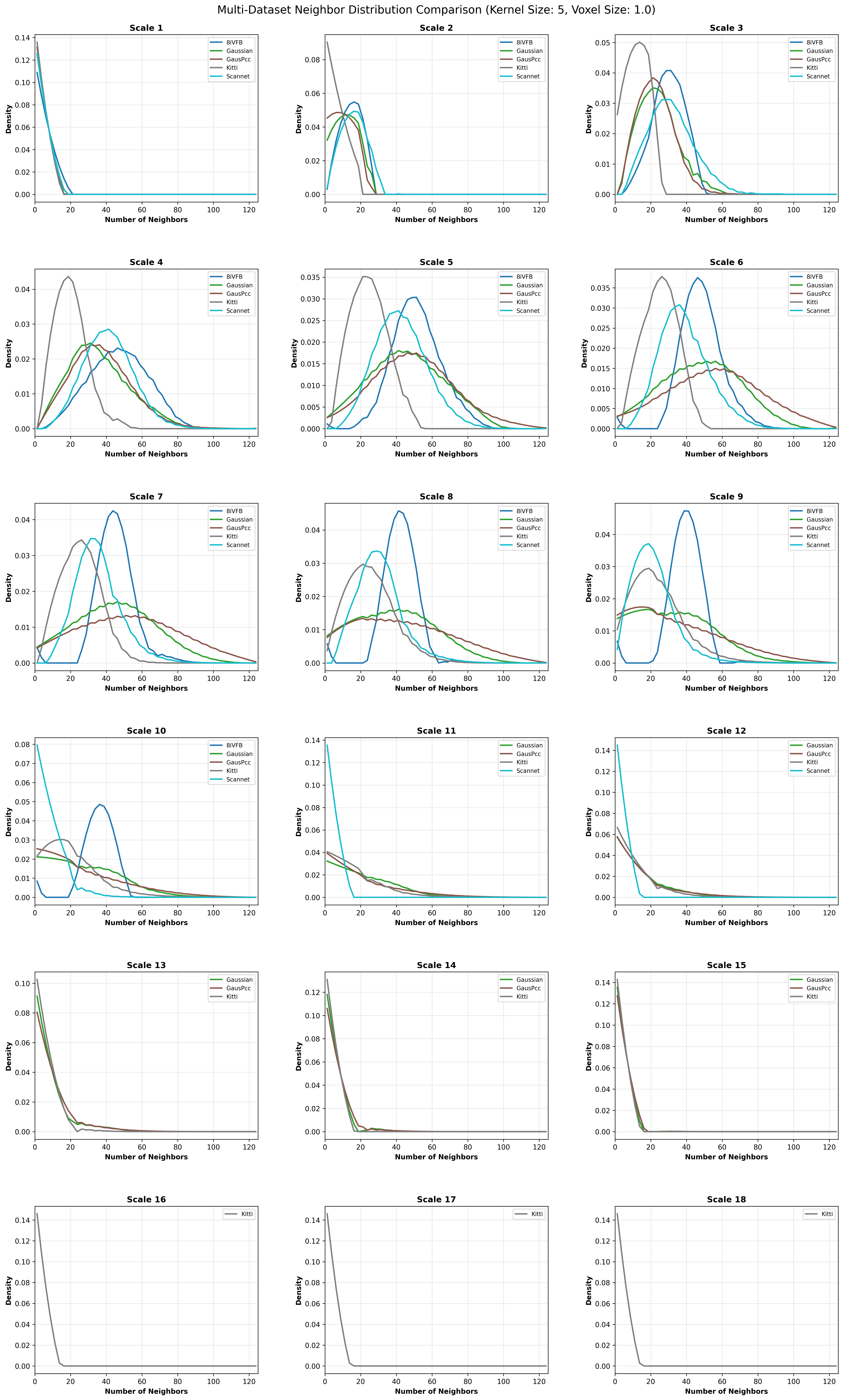}
     \caption{
     Local density comparison of 8iVFB, Gaussian, GausPcc, Kitti, and ScanNet datasets across all scales. Data exceeding the quantization precision of the dataset at higher scales are omitted from the figure.
     }
    \label{fig:all_scale_local_density}
\end{figure}

\begin{figure}
    \centering
    \includegraphics[width=1.0\linewidth]{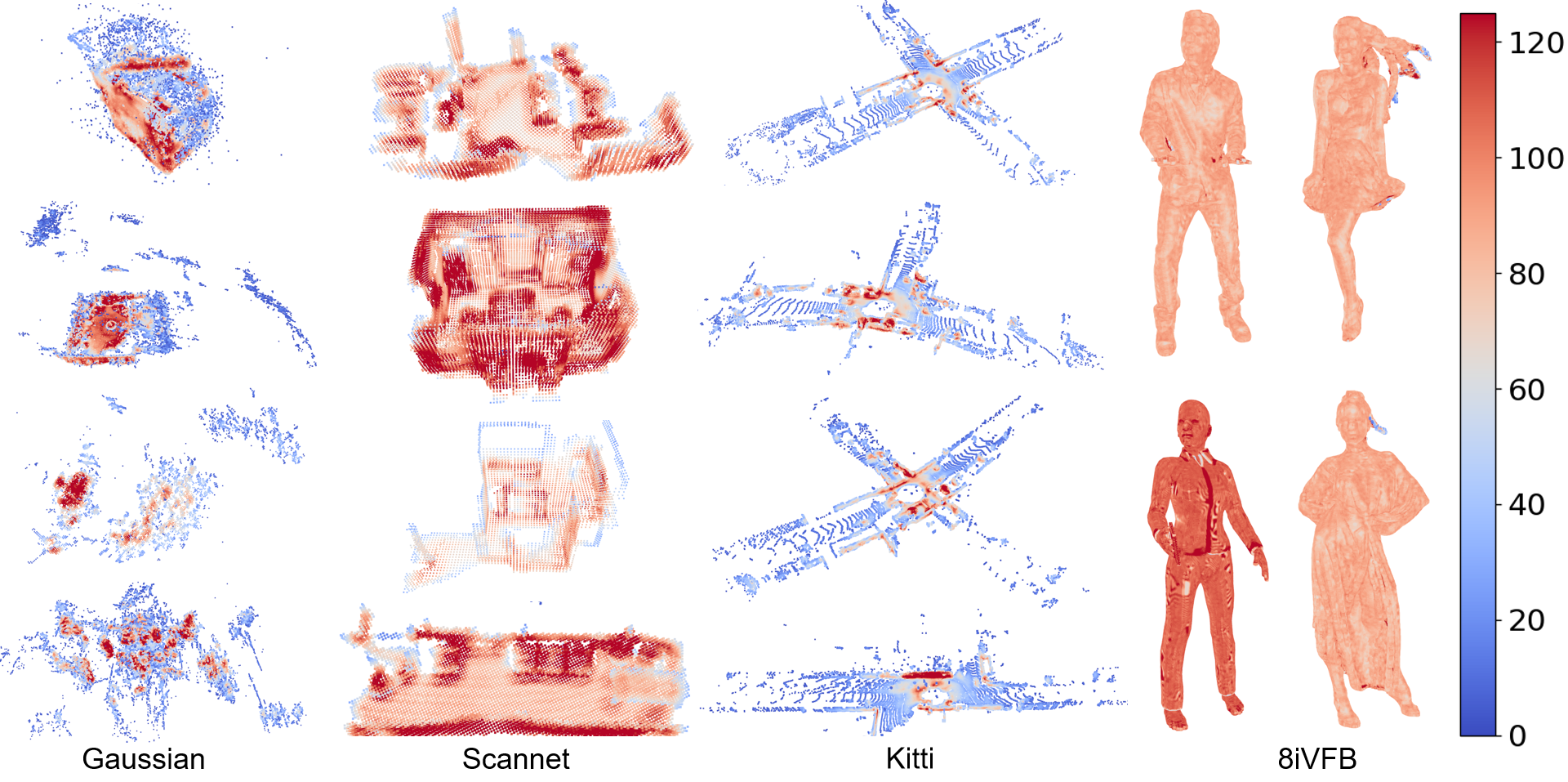}
     \caption{
     Visualization of local density for more samples. All point clouds are visualized at Scale 10, with redder colors indicating higher neighborhood density.
     }
    \label{fig:all_pc_vis}
\end{figure}

\textbf{Further Analysis.}
In Figure \ref{fig:all_scale_local_density}, we present a comparison of local neighbor density across different scales for five datasets, where ``Gaussian'' represents the commonly used evaluation sets in 3D reconstruction and Gaussian compression domains, while ``GausPcc'' refers to the dataset proposed in this paper. At Scales 1-3, the five datasets exhibit similar characteristics. However, at Scales 4-10, Gaussian point clouds demonstrate distribution characteristics that differ significantly from existing datasets, which explains the poor compression efficiency of current AI-based point cloud compression methods. At Scales 11-15, Gaussian point clouds show characteristics similar to the Kitti dataset. From Scales 7-9, Gaussian point clouds generally present a very flat distribution, indicating that the number of neighbors is uniformly distributed between 0 and 124. During training, models typically learn to predict context based on dense or sparse distributions of point clouds, but this particular distribution of Gaussian point clouds makes contextual modeling extremely challenging. Consequently, there is an urgent need for specialized Gaussian point cloud datasets and tailored algorithms. Moreover, across all scales, the distributions of Gaussian and GausPcc are remarkably similar, validating the effectiveness of our proposed dataset. This further confirms that GausPcc is particularly well-suited for training Gaussian point cloud compression methods.

\begin{table}[]
\caption{Kullback-Leibler divergence between Gaussian point clouds and other datasets across scales. Each comparison shows KL divergence from Gaussian to target (first column), target to Gaussian (second column), and symmetric average (third column). The \colorbox[HTML]{FFC7CE}{\textcolor{black}{red}} and \colorbox[HTML]{FFEB9C}{\textcolor{black}{yellow}} highlights indicate the smallest and second-smallest divergence values, revealing datasets most similar to Gaussian Point Clouds distributions. We omit comparative results for Scales 1-3 as their bitstream requirements during compression are negligible.}
\label{table:kl}
\centering
\setlength{\tabcolsep}{5pt} 
\begin{tabular}{c|ccc|ccc|ccc|ccc}
\toprule
\textbf{Scale} & \multicolumn{3}{c|}{\textbf{8iVFB}} & \multicolumn{3}{c|}{\textbf{GausPcc}} & \multicolumn{3}{c|}{\textbf{Kitti}} & \multicolumn{3}{c}{\textbf{Scannet}} \\ \hline
\textbf{4}     & 0.33       & 0.25       & 0.29      & 0.01        & 0.09       & 0.05       & 2.60       & 0.60       & 1.60      & 0.25        & 0.14       & 0.20      \\
\textbf{5}     & 1.24       & 0.22       & 0.73      & 0.03        & 0.17       & 0.10       & 6.37       & 0.78       & 3.57      & 0.45        & 0.15       & 0.30      \\
\textbf{6}     & 2.78       & 0.40       & 1.59      & 0.07        & 0.26       & 0.17       & 8.21       & 0.91       & 4.56      & 0.83        & 0.30       & 0.56      \\
\textbf{7}     & 3.44       & 0.50       & 1.97      & 0.08        & 0.18       & 0.13       & 2.75       & 0.67       & 1.71      & 1.15        & 0.37       & 0.76      \\
\textbf{8}     & 4.77       & 0.66       & 2.71      & 0.06        & 0.17       & 0.12       & 0.63       & 0.35       & 0.49      & 1.15        & 0.35       & 0.75      \\
\textbf{9}     & 5.97       & 0.74       & 3.35      & 0.04        & 0.06       & 0.05       & 0.25       & 0.19       & 0.22      & 0.57        & 0.37       & 0.47      \\
\textbf{10}    & 7.09       & 0.79       & 3.94      & 0.03        & 0.04       & 0.04       & 0.14       & 0.12       & 0.13      & 1.27        & 0.71       & 0.99      \\
\textbf{11}    & N/A        & N/A        & N/A       & 0.03        & 0.04       & 0.04       & 0.06       & 0.05       & 0.05      & 10.44       & 1.03       & 5.73      \\
\textbf{12}    & N/A        & N/A        & N/A       & 0.01        & 0.01       & 0.01       & 0.03       & 0.02       & 0.03      & 6.72        & 0.62       & 3.67      \\
\textbf{13}    & N/A        & N/A        & N/A       & 0.01        & 0.02       & 0.02       & 0.29       & 0.06       & 0.18      & N/A         & N/A        & N/A       \\
\textbf{14}    & N/A        & N/A        & N/A       & 0.03        & 0.24       & 0.13       & 0.50       & 0.05       & 0.27      & N/A         & N/A        & N/A       \\
\textbf{15}    & N/A        & N/A        & N/A       & 0.02        & 0.13       &  0.07       & 0.06       & 0.01       & 0.04      & N/A         & N/A        & N/A       \\ \hline
\textbf{Avg}   & 3.66       & 0.51       & 2.08      & 0.04        & 0.12       & \cellcolor[HTML]{FFC7CE}{0.08}       & 1.82       & 0.32      & \cellcolor[HTML]{FFEB9C}{1.07}      & 2.54        & 0.45       &  1.49      \\  \toprule
\end{tabular}
\end{table}

\subsection{Fractal Dimension Analysis}
\label{Fractal Dimension Analysis}
\begin{figure}
    \centering
    \includegraphics[width=1.0\linewidth]{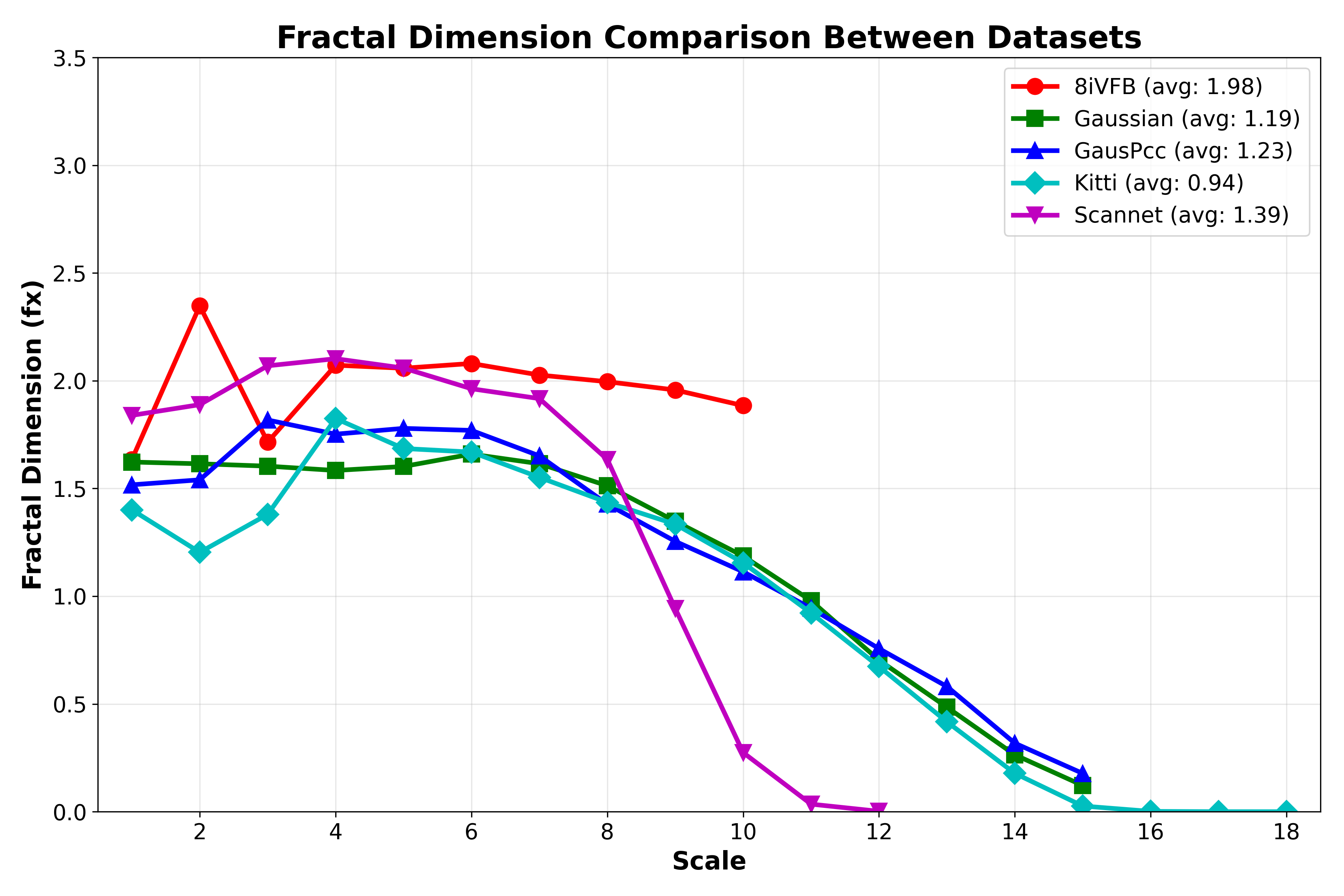}
     \caption{
     Fractal Dimension comparison of 8iVFB, Gaussian, GausPcc, Kitti, and ScanNet datasets across all scales.
     }
    \label{fig:all_datasets_fractal_dimension}
\end{figure}
Fractal dimension is an important metric that describes the degree of self-similarity of geometric objects at different scales, providing a theoretical basis for analyzing the topological properties of point clouds.
The fractal dimension of a point cloud is based on box-counting dimension theory, defined as:
\begin{equation}
D_B = \lim_{\epsilon \to 0} \frac{\log N(\epsilon)}{\log(1/\epsilon)},
\end{equation}
where $N(\epsilon)$ represents the number of boxes with side length $\epsilon$ needed to cover the geometric object.

In practical calculations, we approximate the above limit process using the downsampling sequence $\{\mathcal{X}^{(l)}\}_{l=0}^L$, where the voxel size doubles with each downsampling: $\epsilon_l = \epsilon_0 \cdot 2^l$, with $\epsilon_0$ being the initial voxel size and $l$ the scale index.

For each scale $l$, we record the number of voxels $N_l = |\mathcal{X}^{(l)}|$. According to fractal theory, under ideal conditions:
\begin{equation}
N_l \approx C \cdot \epsilon_l^{-D_B},
\end{equation}
where $C$ is a constant and $D_B$ is the fractal dimension.
For two adjacent scales $l$ and $l+1$, taking the logarithm and applying differentiation yields the fractal dimension estimate between them:
\begin{equation}
f_x(l, l+1) = \frac{\log N_l - \log N_{l+1}}{\log \epsilon_{l+1} - \log \epsilon_l}.
\end{equation}
Since the voxel size of adjacent scales has a 2-fold relationship: $\epsilon_{l+1} = 2\epsilon_l$, the above equation simplifies to:
\begin{equation}
f_x(l, l+1) = \frac{\log N_l - \log N_{l+1}}{\log 2} = \log_2 \frac{N_l}{N_{l+1}}.
\label{fractal dimension cal}
\end{equation}

\textbf{Further Analysis.}
As illustrated in Figure \ref{fig:all_datasets_fractal_dimension}, we present a comparison of fractal dimensions across five datasets. Notably, in terms of overall density variation, Gaussian point clouds exhibit characteristics similar to those of Kitti LiDAR point clouds, corresponding with the visualization in Figure \ref{fig:all_pc_vis}.

\section{Limitations, Future Work and Broader impacts}
\subsection{Limitations}
\label{sub:limit}
The proposed method inevitably introduces temporal latency during encoding and decoding processes, and its encoding efficiency may be compromised when the scene contains numerous duplicate coordinate points.
\subsection{Future Work}
Future work will focus on two primary directions: First, developing a Gaussian point cloud attribute compression (GausPcac) scheme by training generalizable neural compressors using the GausPcc-1K dataset. This would enable GausPcc (GausPcgc + GausPcac) to directly compress trained scenes without requiring retraining as in Cat-3DGS \cite{zhancat}. Second, improving encoding efficiency for scenes with numerous duplicate points, for instance by encoding the point repetition count as a point cloud attribute and restoring it during decoding.
\subsection{Broader impacts}
\label{sub:Broader}
Our work on 3D scene compression representation offers several positive societal impacts: (1) By significantly reducing storage and transmission costs for 3D content, we make high-quality 3D scenes more accessible to resource-constrained applications and users; (2) This improved accessibility enables broader adoption in educational settings, where 3D visualization can enhance learning experiences; (3) For virtual reality applications, our compression technique enables more efficient streaming and rendering of complex 3D environments, making VR more accessible to broader audiences.

However, we acknowledge potential negative implications: the ability to efficiently compress and transmit realistic 3D scenes could be misused to create and distribute misleading or deceptive 3D content, particularly in the context of deepfakes. Nevertheless, numerous technologies \cite{xu2024fakeshield,zhang2024gs} are already being developed to mitigate the impact of such synthetic media.

\end{document}